\documentclass[reprint,
superscriptaddress,
aps, pra
]{revtex4-2}

\usepackage{graphicx}% Include figure files
\usepackage{dcolumn}% Align table columns on decimal point
\usepackage{bm}% bold math
\usepackage{parskip}
\usepackage{graphicx}
\usepackage{braket}
\usepackage{tabularx}
\usepackage{upgreek}
\let\PhysLett\pl
\let\pl\relax
\usepackage{unitsdef}
\let\pl\PhysLett

\usepackage{xcolor}
\usepackage{amsmath}
\usepackage{hyperref}
\usepackage{url}

\usepackage[english]{babel}
\makeatletter
\def\l@en{\l@english}
\makeatother

\makeatletter
\def\l@eng{\l@english}
\makeatother
\begin{document}

\preprint{APS/123-QED}

\title{Spin Readout in a 22~nm Node Integrated Circuit}
%\title{Spin readout in a deep sub-micron integrated circuit}
%\title{Spin Readout of a Quantum Dot in a 22nm FDSOI Integrated Circuit}

\author{Isobel~C.~Clarke}
\email{isobel@quantummotion.tech}
\affiliation{Quantum Motion, 9 Sterling Way, London, N7 9HJ, United Kingdom}
\affiliation{London Centre for Nanotechnology, UCL, London, WC1H 0AH, United Kingdom} %Department of EE? 
\author{Virginia Ciriano-Tejel}
\affiliation{Quantum Motion, 9 Sterling Way, London, N7 9HJ, United Kingdom}
\author{David J. Ibberson}
\affiliation{Quantum Motion, 9 Sterling Way, London, N7 9HJ, United Kingdom}
\author{Grayson~M.~Noah}
\affiliation{Quantum Motion, 9 Sterling Way, London, N7 9HJ, United Kingdom}
\author{Thomas H. Swift}
\affiliation{Quantum Motion, 9 Sterling Way, London, N7 9HJ, United Kingdom}
\author{Mark~A.~I.~Johnson}
\affiliation{Quantum Motion, 9 Sterling Way, London, N7 9HJ, United Kingdom}
\author{Ross C. C. Leon}
\affiliation{Quantum Motion, 9 Sterling Way, London, N7 9HJ, United Kingdom}
\author{Alberto~Gomez-Saiz}
\affiliation{Quantum Motion, 9 Sterling Way, London, N7 9HJ, United Kingdom}
\affiliation{Department of Electrical and Electronic Engineering, Imperial College London, London SW7 2AZ, United Kingdom}
\author{John~J.~L.~Morton}
\email{john@quantummotion.tech}
\affiliation{Quantum Motion, 9 Sterling Way, London, N7 9HJ, United Kingdom}
\affiliation{London Centre for Nanotechnology, UCL, London, WC1H 0AH, United Kingdom}
\author{M. Fernando Gonzalez-Zalba}
\email{fernando@quantummotion.tech}
\affiliation{Quantum Motion, 9 Sterling Way, London, N7 9HJ, United Kingdom}
\affiliation{CIC nanoGUNE Consolider, Tolosa Hiribidea 76, E-20018 Donostia-San Sebastian, Spain}
\affiliation{IKERBASQUE, Basque Foundation for Science, E-48011 Bilbao, Spain}

\date{\today}

\begin{abstract}
% Building a useful quantum computing system will require millions of physical qubits. 
% which can be addressed with control and readout electronics at cryo/on-chip. 
% modern CMOS .. and present the possibility of integrating spin qubits with classical electronics.  
% %Silicon-based devices are a promising platform to build the million of qubits required for a useful QCS. 
% Current silicon-based devices fabricated using academic and industry-compatible processes have grown to a few qubits however further scaling limited. 
% %and perform simple error-corrected quantum algorithms. 
% Here, we utilize the ubiquious silicon Integrated Circuit (IC) to offer a solution.
% %ever-present/universal 
% %Utilizing the ever-present/universal silicon integrated circuit (IC) offers a solution. 
% The first step to forming spin-based qubits in this platform is to perform a spin state measurement. Hence, we present single-shot readout of spins in quantum dot (QD) devices hosted on an IC fabricated using and addressed on-chip industry-standard 22nm-node technology. We use a ramped energy-selective measurement to perform spin-to-charge conversion, detected using a radio-frequency single-electron transistor (rfSET). Similar readout fidelities above 80\% and milli-second spin relaxation times in two identical devices demonstrates the reproducibility of the QD unit cell and the promise of scaling the device across the IC.
%commerical?

Constructing a quantum computer capable of broad and important applications is likely to require millions of addressable physical qubits, posing the challenge of large-scale integration of quantum systems with classical electronics.
Fully depleted silicon-on-insulator CMOS technology has been used to develop a range of  cryogenic electronic components for the control and readout of different qubit modalities interfaced on separate chips. However, recent measurements of quantum dots on this technology raise the tantalising prospect of realising control electronics and spin qubits on the same manufacturing platform, within a single integrated circuit (IC).
Here, we demonstrate single-shot spin readout in addressable quantum dot devices within an IC fabricated using industry-standard 22~nm fully depleted silicon-on-insulator technology. We achieve spin-to-charge conversion via a ramped energy-selective measurement, detected using a radio-frequency single-electron transistor and addressed by on-chip cryogenic electronics. The observation of consistent readout visibilities exceeding 90\% and millisecond spin relaxation times in two nominally identical devices within the addressable array supports the reproducibility of the unit cell. The successful observation of spin readout using this CMOS process marks a key step towards realising highly scalable and integrated spin qubits. 

%Building a useful quantum computing system will require millions of physical qubits addressed via cryogenic circuitry.
%Modern silicon Integrated Circuit (IC) technology offers the possibility of integrating spin-based qubits and the required classical electronics on the same chip. The first step to forming a spin qubit in this platform is to perform a spin state measurement. Here, we present single-shot readout of spins in quantum dot (QD) devices hosted and addressed on an IC fabricated using industry-standard 22~nm-node technology. We use a ramped energy-selective measurement to perform spin-to-charge conversion, detected using a radio-frequency single-electron transistor (rfSET). Similar readout fidelities above 80\% and milli-second spin relaxation times in two identical devices demonstrates the reproducibility of the QD unit cell and the promise of scaling the device across the IC.

\end{abstract}

\maketitle 

\section{Introduction}
% Quantum processor units (QPU) are predicted to need millions of physical qubits to perform 
% %useful/relevent/
% practical fault-tolerant algorithms. %[ref?]. 
% Spin qubits hosted on silicon quantum dot (QD) devices are ideal candidates to build a scalable QPU. %This is due to/
% Benefits include their small footprint, qubit preparation \cite{Yoneda2020, PRXQuantum.3.010352} and control fidelities \cite{Mills2022,Yoneda2018,Noiri2022} above the 99\% error-rate threshold \cite{PhysRevA.86.032324}, and the manufacturing potential of leveraging decades %/$\sim$65 years  
% of advancements in metal-oxide-semiconductor (MOS) IC technology. Moreover, recent developments in cryo-CMOS circuitry holds promise for building an interface between the silicon QPU and classical electronics at milli-kelvin temperatures \cite{Ruffino2022, Pauka2021,Bartee2025}, 
% a requirement for future large-scale systems \cite{fernando2021}.
% Forming a qubit first requires demonstrating single-shot spin readout of a single electron. This involves projecting the electron spin to a charge state, detected using a nearby charge sensor. A commonly used spin-to-charge conversion technique is called energy-selective readout (ERO) \cite{elzerman_2004}. 
% To date, spin readout has been performed in silicon MOS QD devices fabricated in academic-style cleanrooms \cite{Veldhorst2014,Fogarty2018,Urdampilleta2019} and using industrially-compatible processes \cite{Zwerver2022,Virginia2021,steinacker2024}.  
In the prototype quantum computing systems which exist today~\cite{Madsen2022, Kim2023, Bluvstein2024, Xu2025}, the quantum processing unit (QPU) and the classical control electronics are physically separated, for example by a cryostat or vacuum chamber. While such a `modular' approach may suffice at the level of tens to hundreds of qubits, scaling to the millions of physical qubits needed to perform practical fault-tolerant algorithms~\cite{Fowler2012} requires advanced integration of qubits with the electronics for addressing, control and readout~\cite{fernando2021}.
There are two primary approaches to such integration: (i) \emph{heterogeneous integration}, in which QPUs fabricated using a bespoke qubit process are interfaced with chips containing analogue and digital electronics, e.g.\ via bondwires or microbumps~\cite{Xue2021,Paradkar2025FlipChipIndium,Bartee2025}; 
and (ii) \emph{homogeneous integration}, in which qubits have been co-fabricated alongside the classical control and readout electronics using the same industrial technology platform on a single integrated circuit~\cite{King2023QuantumCritical5000}.
%~\cite{Schaal2019, Guevel2020, Ruffino2022, amitonov2024}. 
Both approaches bring benefits in terms of scaling such as reduced footprint and better signal integrity, but have the challenge of operating the control electronics in the same environmental conditions (e.g.\ temperature) as the QPU, while avoiding heat or noise affecting QPU performance~\cite{reilly2019challengesscalingupcontrolinterface}. Homogeneous integration in particular offers major potential advantages in scalability and performance through the removal of interconnects, however, it carries the significant challenge of realising qubits on a platform capable of advanced, low-power electronics.
%However, a demonstration that integrates spin-based quantum functionality with classical electronics on a single chip has yet to be realized.

%CryoCMOS used to control qubits:
%~\cite{Xue2021,Pauka2021,Bartee2025

%CryoCMOS + QDs off-chip
%~\cite{Schaal2019}. 
%CryoCMOS + QDs on-chip
%~\cite{Guevel2020, Ruffino2022, amitonov2024, Thomas2025}

% put figure here for top of next page 
\begin{figure*}[t]
\includegraphics[width=\textwidth]{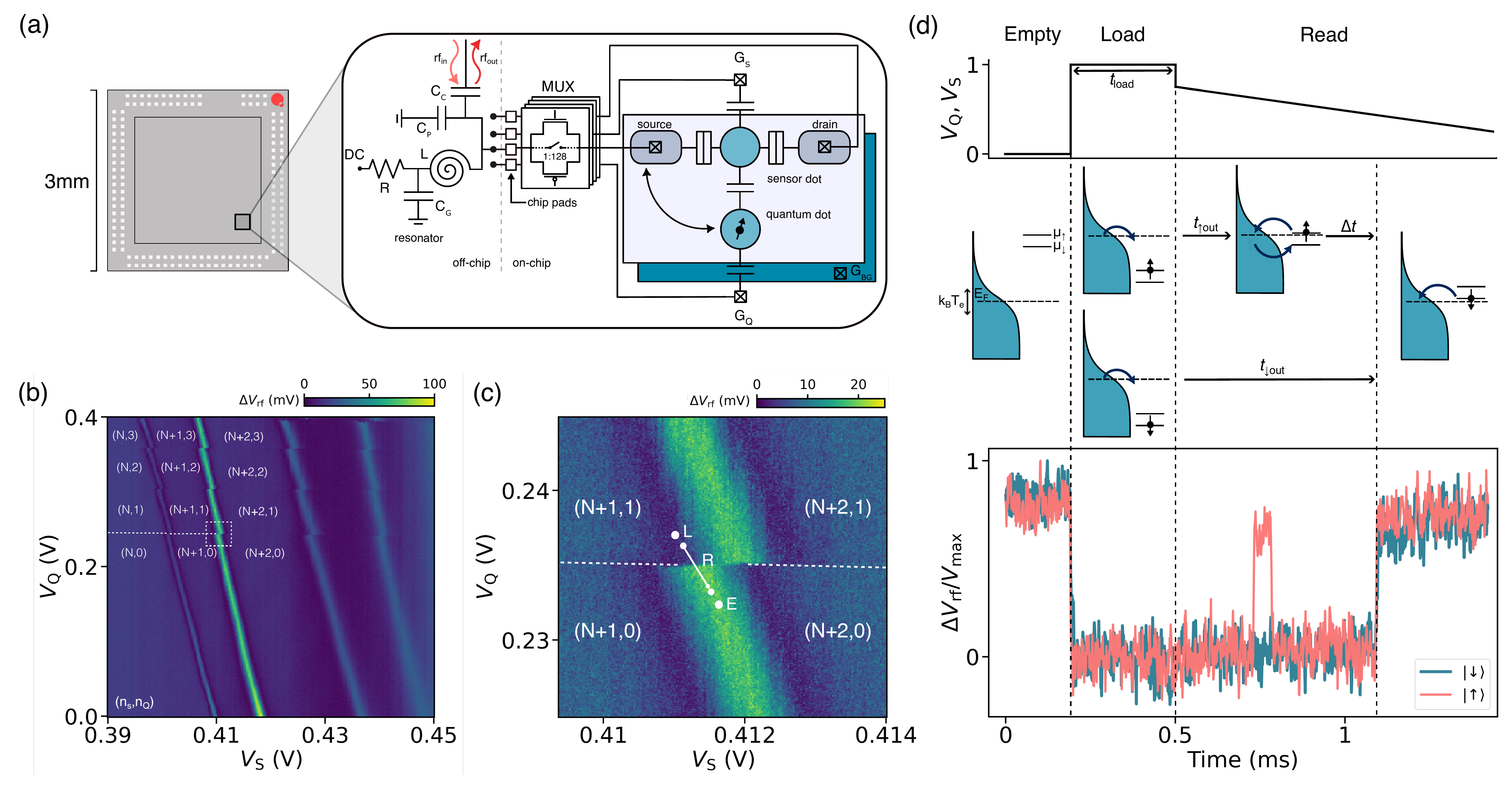}
  \caption{\textbf{Device schematic and spin readout protocol.}
  a) 22~nm FDSOI IC hosting 128 unit cells formed by the QD structures and the integrated multiplexing circuitry. Zoom in of the unit cells used in this experiment containing a few-electron QD with a rfSET charge sensor in parallel. The off-chip $LC$ resonant circuit contains a superconducting spiral inductor, $L =$ 18.7 nH, coupled to the transmission line via coupling capacitor $C_C =$ 0.9 pF and a parasitic capacitance of $C_P \sim$ 2.5 pF. A bias tee containing a resistor $R$ and capacitor $C_G$ is used to apply a dc source voltage and create a good rf ground. 
  b) Charge stability diagram of the sensor-QD system in device A labelled with the electron occupation on each dot. The dotted square indicates the chosen transition to perform spin readout. Device B shows a similar stability map and single-electron charge transition chosen for spin readout. 
  c) Charge transition $(n_\text{S},n_\text{Q}) = (N+1, 1) \rightarrow (N+2, 0)$ in b). Pulse sequence with Empty-Load-Read stages is applied to cycle the electron occupation of the qubit dot from 0 to 1. A random electron spin state is initialised in the load stage. 
  d) Voltage pulse sequence applied to gates $G_\text{Q}$ and $G_\text{S}$ and the subsequent %resulting
  response of the qubit dot electrochemical potentials, $\mu_{\uparrow/\downarrow}$, and the single-shot charge sensor signal. A `blip' in signal before a certain time threshold indicates a $\ket{\uparrow}$ electron loaded onto the dot.
  }
  \label{fig:device setup}
\end{figure*}

Given the dominance of silicon-based technology in conventional digital and analogue electronics, spin qubits confined in silicon quantum dots (QDs) represent particularly compelling candidates for building a homogeneously integrated quantum computing system~\cite{fernando2021}. As a materials platform, spin qubits in silicon have shown preparation, control and readout fidelities above the 99\% threshold of the surface code~\cite{Yoneda2020, PRXQuantum.3.010352, Mills2022,Yoneda2018,Noiri2022} and have been manufactured using bespoke 300~mm wafer industrial grade processes~\cite{steinacker2024,Zwerver2022,chittockwood2025,Laine2025,hamonic2025}, while cryogenic CMOS electronics have been developed for the control, operation and readout of spin-based quantum processors~\cite{Xue2021,Pauka2021,Ibberson2024,Bartee2025}. 
%and the potential to leverage industrial complementary metal-oxide-semiconductor (CMOS) IC technology to manufacture both the QPU and the classical control circuitry~.
%Integration at cryogenic temperatures offers many benefits in terms of scaling such as reduced footprint, better signal integrity, and efficient input/output management if power dissipation can be managed~\cite{reilly2019challengesscalingupcontrolinterface}. 
Silicon spin qubits also offer the considerable advantage of operating at elevated cryogenic temperatures ($\approx 1$ ~K)~\cite{Huang2024}, where substantially higher cooling power is available compared to qubit systems that operate around 10~mK, increasing the scope of integration with dissipative control electronics.  
As an illustration of the potential feasibility of homogeneously integrating spin qubits with on-chip control, arrays of addressable QDs have been fabricated alongside cryogenic electronics on industrial CMOS processes~\cite{Guevel2020, Ruffino2022, amitonov2024}. Recently, a GlobalFoundries 22FDX silicon-on-insulator process has been used to manufacture an addressable array of over a thousand QD devices integrated with transistor digital logic on a single 9 mm$^2$ silicon chip~\cite{Thomas2025}. However, to realise and study qubits in this technology it is necessary to demonstrate single-electron occupation of such QDs and isolate the spin degree of freedom.%., performing single-shot spin readout.
%
%In terms of integration strategies, the field has primarily advanced along two paths: (i) \textbf{heterogeneous integration}, which combines QPUs fabricated in academic-style cleanrooms with industry-standard chips that perform classical logic~\cite{Xue2021,Pauka2021,Bartee2025}, and (ii) \textbf{homogeneous integration}, in which QD-based devices have been co-fabricated with classical control and readout electronics using the same industrial technology platform~\cite{Schaal2019, Guevel2020, Ruffino2022, amitonov2024}. However, a demonstration that integrates spin-based quantum functionality with classical electronics on a single chip has yet to be realized.
%
%Forming a qubit first requires demonstrating single-shot spin readout of a single electron. This involves projecting the electron spin to a charge state, detected using a nearby charge sensor. A commonly used spin-to-charge conversion technique is called energy-selective readout (ERO)~\cite{elzerman_2004}.  To date, spin readout has been performed in silicon MOS QD devices fabricated in academic-style cleanrooms~\cite{Veldhorst2014,Fogarty2018,Urdampilleta2019} and using industrially-compatible processes~\cite{Virginia2021,Zwerver2022,chittockwood2025,steinacker2024, Laine2025}.
%

In this article, we present single-shot spin readout of QD devices hosted and addressed via a digitally-controlled analog multiplexer (MUX) on an IC fabricated using industry-standard 22~nm fully depleted silicon-on-insulator (FDSOI) technology (GlobalFoundries 22FDX). 
%Forming a qubit first requires demonstrating single-shot spin readout of a single electron. This involves projecting the electron spin to a charge state, detected using a nearby charge sensor. A commonly used spin-to-charge conversion technique is called energy-selective readout (ERO)~\cite{elzerman_2004}.  To date, spin readout has been performed in silicon MOS QD devices fabricated in academic-style cleanrooms~\cite{Veldhorst2014,Fogarty2018,Urdampilleta2019} and using industrially-compatible processes~\cite{Virginia2021,Zwerver2022,chittockwood2025,steinacker2024, Laine2025}.
We develop a unit cell consisting of a radio-frequency single-electron transistor and a few-electron QD. We use a ramped energy-selective readout~\cite{elzerman_2004,Keith2022} on two nominally identical devices within the same IC, selected by row-column addressing using an on-chip MUX~\cite{Thomas2025}. Finally, we analyse the spin readout visibility of the method and outline future directions for the development of this approach.  %An alternative Classifier analysis method to the commonly used threshold method is presented to improve readout fidelity at low magnetic fields, $B < 2$~T, in the high electron temperature regime. %In this way, we demonstrate substantial steps in forming a qubit in industrially-fabricated ICs.  

%We use GlobalFoundries' 22~nm FDSOI process of GlobalFoundries, called 22FDX, capable of manufacturing billions of qubit-hosting transistors integrated with the classical electronics on a single 3~mm square chip. We implement a ramped energy-selective readout method called ramped spin measurement (RSM) \cite{Keith2022} on two nominally identical devices within the same IC, selected by row-column addressing using an on-chip multiplexer (MUX)~\cite{Thomas2025}. An alternative Classifier analysis method to the commonly used threshold method is presented to improve readout fidelity at low magnetic fields, $B < 2$~T, in the high electron temperature regime. In this way, we demonstrate substantial steps in forming a qubit in industrially-fabricated ICs. 

\section{Multiplexed spin readout unit cell}

%We use an integrated circuit consisting of a spin readout unit cell and a cryogenic multiplexer

The circuit used for spin readout is part of a 3~x~3~mm silicon IC fabricated using GlobalFoundries 22-nm FDSOI technology; see Fig.~\ref{fig:device setup}a. The IC contains a one-to-128 multiplexer with a 7-bit digital select bus that enables the sequential addressing of individual devices among an array of 128 different device variants (see Methods). The subset of devices used for spin readout consists of a single-electron transistor (SET) charge sensor capacitively coupled to a few-electron QD whose physical dimensions are varied across the array. The MUX enables rapid device testing to find the optimal device dimensions. The unit cell for spin readout consists of two 28-nm-long gates placed face-to-face on the top side of the same ultra-thin silicon body. The sensor gate $G_\text{S}$ primarily controls the current flow through the SET (which is tunnel coupled to highly doped source and drain contacts), whereas the qubit gate $G_\text{Q}$ controls primarily the electrostatic potential of the QD. In addition, the structure has a global back gate $G_\text{BG}$, implemented by applying an electrostatic potential to the N-well under the device which is isolated from the 6-nm ultra-thin-body by a 20-nm thick buried oxide~\cite{Wang2022}. Finally, we implement a top gate $G_\text{TG}$, consisting of a metal electrode from the lower interconnect layer of the technology to allow further tuning.

To increase the bandwidth of the SET and implement fast charge sensing, we embed the sensor in an $LC$ radio-frequency (rf) matching network by connecting an off-chip superconducting inductor $L$ and a surface mount coupling capacitor $C_\text{C}$ to the source input of the MUX. Such an arrangement enables time-division multiplexing of the desired cells. We evaluate the performance of the rfSET using the minimum integration time to achieve a signal-to-noise ratio (SNR) of 1, $t_\text{min}$~\cite{Vigneau2023}. We measure $t_\text{min} = 9$~ns at zero magnetic field which will later comfortably enable single-shot spin state readout via spin-dependent tunneling. While the resonant $LC$ circuit used here is off-chip, demonstrated compact superinductors in the same technology could be used to integrate the resonator in the future~\cite{swift2025}. 

 %only $\sim$ 10 times larger than state-of-the-the-art SET sensors used for single-shot readout \cite{Keith2019}. % 625ps

We explore the charge stability diagram of the system by monitoring the reflected voltage from the matching network as a function of sensor and qubit gate potentials; see Fig.~\ref{fig:device setup}b. Shifts in the position of the quasi-vertical SET's Coulomb oscillations correspond to an electron tunneling into the QD from the reservoirs. The absence of shifts below $V_\text{Q}=0.23$~V indicates that the QD can be completely depleted of electrons. In the last electron regime ($n_\text{Q}=1$), tunneling can be spin-dependent when a magnetic field is applied to lift the spin degeneracy of electron states $\ket{\uparrow}$ and $\ket{\downarrow}$. Here, we apply the field in the plane of the device. We measure at the $(n_\text{S},n_\text{Q}) = (N+1, 1) \rightarrow (N+2, 0)$ charge transition as the rfSET signal is largest. We perform a ramped spin measurement (RSM) using three-stage voltage pulse sequence with phases Empty (E), Load (L) and Read (R). We apply the potentials to the sensor and qubit dot gates simultaneously to move diagonally across the charge sensing transition in Fig.~\ref{fig:device setup}c. We map the spin states to a charge state by measuring the sensor response during the pulse sequence. The single-shot traces show maximum signal when the dot is empty and low background signal when occupied (Fig.~\ref{fig:device setup}d).

\section{Ramped Spin Readout}

\begin{figure}[t!]
    \includegraphics[width=\columnwidth]{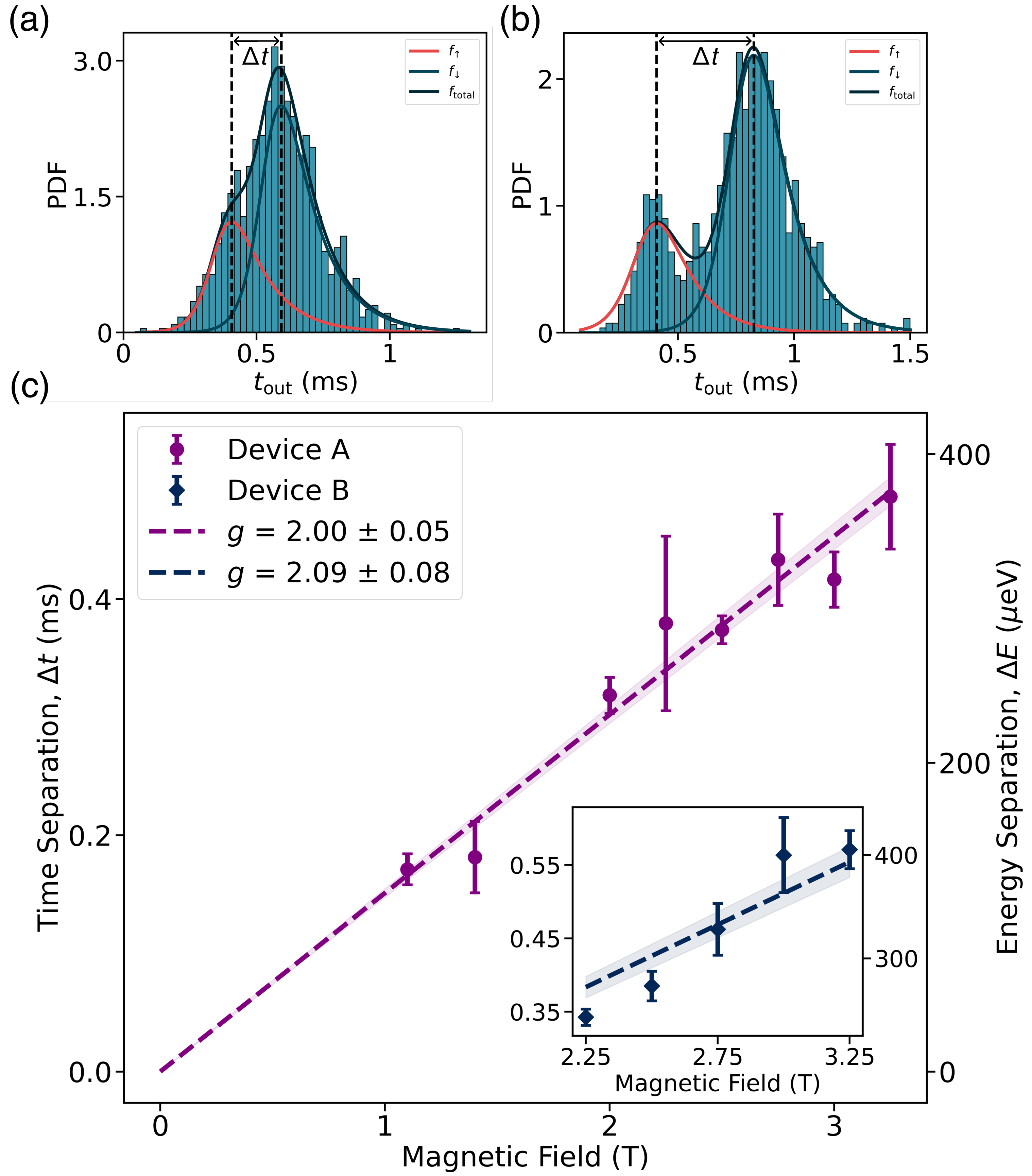}
    \caption{\textbf{Single-shot spin readout.} Histograms of the first tunneling out event, $t_{\text{out}}$, for 1000 single-shot traces at $B = 1.4$~T (a) and $B = 2.5$~T (b). Both distributions are fit to a generalized probability density function, $f_{\text{total}}(t)$, and the resulting individual spin state probability density functions, $f_{\uparrow / \downarrow}(t)$, are indicated. Dashed lines show the center of each spin state peak at time when $df_{\uparrow / \downarrow}(t)/dt = 0$.
    c) Time separation between spin state tunneling out events as a function of magnetic field for both devices. 
    %Straight line fit predicts electron spin degeneracy at $B = 0$~T, within experimental error. 
    Shaded region shows bounds of the fit. Error bars plotted are given in terms of energy. %Converting into energy, the peak separation becomes proportional to the Zeeman energy at each field. 
    }
    \label{fig:peak sep}
\end{figure}

For spin-to-charge conversion, we use a form of energy-selective spin readout~\cite{elzerman_2004, Morello2010} with a ramped voltage applied during the `read' phase to improve robustness to noise~\cite{Keith2022}.
%ramped spin measurement (RSM) approach which differs from conventional energy-selective spin readout only in the `Read' stage. In the ERO Read stage, a constant voltage pulse tunes the spin levels to straddle the Fermi level, $E_F$~\cite{elzerman_2004, Morello2010}. 
%In this configuration, a $\ket{\uparrow}$ electron may tunnel out of the QD and be replaced by a $\ket{\downarrow}$ electron creating a `blip’ in sensor signal, while a $\ket{\downarrow}$ electron will remain on the QD, producing no `blip'. Hence, the presence or absence of a `blip’ during the Read phase labels the electron spin state. However, noise during the Read phase caused, for example, by charge noise or bias voltage fluctuations may result in labeling errors. 
%RSM improves upon the noise sensitivity of ERO by ramping the electrochemical levels of the QD through $E_F$ during the Read phase~\cite{Keith2022}. 
The pulse sequence used is shown in Fig.~\ref{fig:device setup}d, beginning by making $V_Q$ sufficiently low to empty the QD, and then pulsing it high to load a single electron with a random spin state. During the `read' phase, $V_Q$ is gradually ramped down, enabling the discrimination of the spin-up  $\ket{\uparrow}$ and spin-down $\ket{\downarrow}$ states of the electron due to their Zeeman energy difference in the applied magnetic field. If a higher energy $\ket{\uparrow}$ electron occupies the QD, it will tunnel out after $t_{\uparrow \text{out}}$, and be replaced by a $\ket{\downarrow}$ electron. The momentary change in QD charge state is registered as a `blip' by the charge sensor. As $V_Q$ continues to ramp down, after some further time $\Delta t$, the $\ket{\downarrow}$ level moves above $E_F$ and tunnels out at $t_{\downarrow \text{out}}$ producing a stepwise change of the sensor response (see Fig.~\ref{fig:device setup}d). In the case of an initial $\ket{\downarrow}$, only the stepwise response is observed. Such a RSM hence provides information to enable spin labelling in the time domain by recording the time at which the first tunneling event occurs~\cite{Keith2022}. On average, it is expected that a $\ket{\uparrow}$ electron will tunnel before a $\ket{\downarrow}$ electron. To setup RSM, we first tune the electron tunnel rate using the front and back gates and then tune the ramp rate so it is fast enough to minimise spin mapping errors produced by relaxation or thermal processes~\cite{Oakes2023} but slow enough to observe a time difference between spin orientations, as we elaborate below.

To determine the spin state from a single-shot trace, we utilize the time-domain information. We implement a threshold method that determines the first time the sensor signal exceeds a voltage threshold $t_{\text{out}}$, i.e. the first time a spin tunnels out of the QD~\cite{Keith2022}. Applying the method to repeated experimental shots reveals a bimodal distribution; see Fig.~\ref{fig:peak sep}a. The earlier peak in the distribution corresponds to $\ket{\uparrow}$ events whereas the later peak corresponds to $\ket{\downarrow}$ events. A further threshold in the time domain at the point of minimum overlap, enables labelling the spin state depending on whether  $t_{\text{out}}$ occurs before or after the threshold. The time separation between distributions can be characterized here by the parameter $\Delta t$, which we obtain by fitting the probability density functions of the spin $\ket{\uparrow}$/$\ket{\downarrow}$ tunneling processes as we shall see later in Section~\ref{sec:fidelity}. At higher magnetic field, the distributions are temporally more separated, see Fig.~\ref{fig:peak sep}b.

%To determine the spin state from a single-shot trace, we first utilize the time-domain information. We implement a standard threshold method that determines the first time the sensor signal exceeds a voltage threshold $t_{\text{out}}$, i.e. the first time a spin tunnels out of the QD~\cite{Keith2022}. Applying the method to repeated experimental shots reveals a bimodal distribution; see Fig. \ref{fig:peak sep}a. The earlier peak in the distribution corresponds to $\ket{\uparrow}$ events whereas the later peak corresponds to $\ket{\downarrow}$ events. A further threshold in the time domain at the point of minimum overlap, enables labelling the spin state depending on whether  $t_{\text{out}}$ occurs before or after the threshold. 

%To improve upon this standard paradigm, we introduce a classification method that utilizes both the time-domain and existence(or absence) of a signal "blip". Particularly, the \textit{Classifier} uses a penalized change point detection method (See Methods). Labeling spin tunneling events relative to a reference point unique to each trace makes the method more robust to low frequency detuning fluctuations of the QD than the static time threshold in the previous method. We plot the normalised distributions $t_{\uparrow \text{out}}$ and $t_{\downarrow \text{out}}$ in Fig. \ref{fig:peak sep}b.

The increased temporal separation at higher magnetic field is indicative of the physical origin of the energy separation between $\ket{\uparrow}$ and $\ket{\downarrow}$ states, i.e. the Zeeman energy $E_\text{Z}=g\mu_\text{B}B$, which is proportional to the magnetic field. Here, $g$ is the electron g-factor which, for an electron in silicon, is $\approx~2$. We confirm this hypothesis by plotting $\Delta t$ versus the applied magnetic field and observe a linear dependence, see Fig.~\ref{fig:peak sep}c. Converting $\Delta t$ into energy using knowledge of the gate lever arms of the device and the gate voltage ramp rates (see Methods), we extract an electron g-factor of 2.00 $\pm$ 0.05 for device A. %and an intercept $\Delta E$ = 4 $\pm$ 25~$\mu$eV.
We perform the same analysis for a second nominally identical device and obtain $g=2.09 \pm 0.08$. The magnetic field dependence of the distributions confirms that electron tunneling is spin-dependent and represents the first evidence of spin readout on a silicon nanodevice fabricated using industry-standard techniques.  %, lower than expected? 

\section{Spin Relaxation}
\begin{figure}[t!]
    \includegraphics[width=\columnwidth]{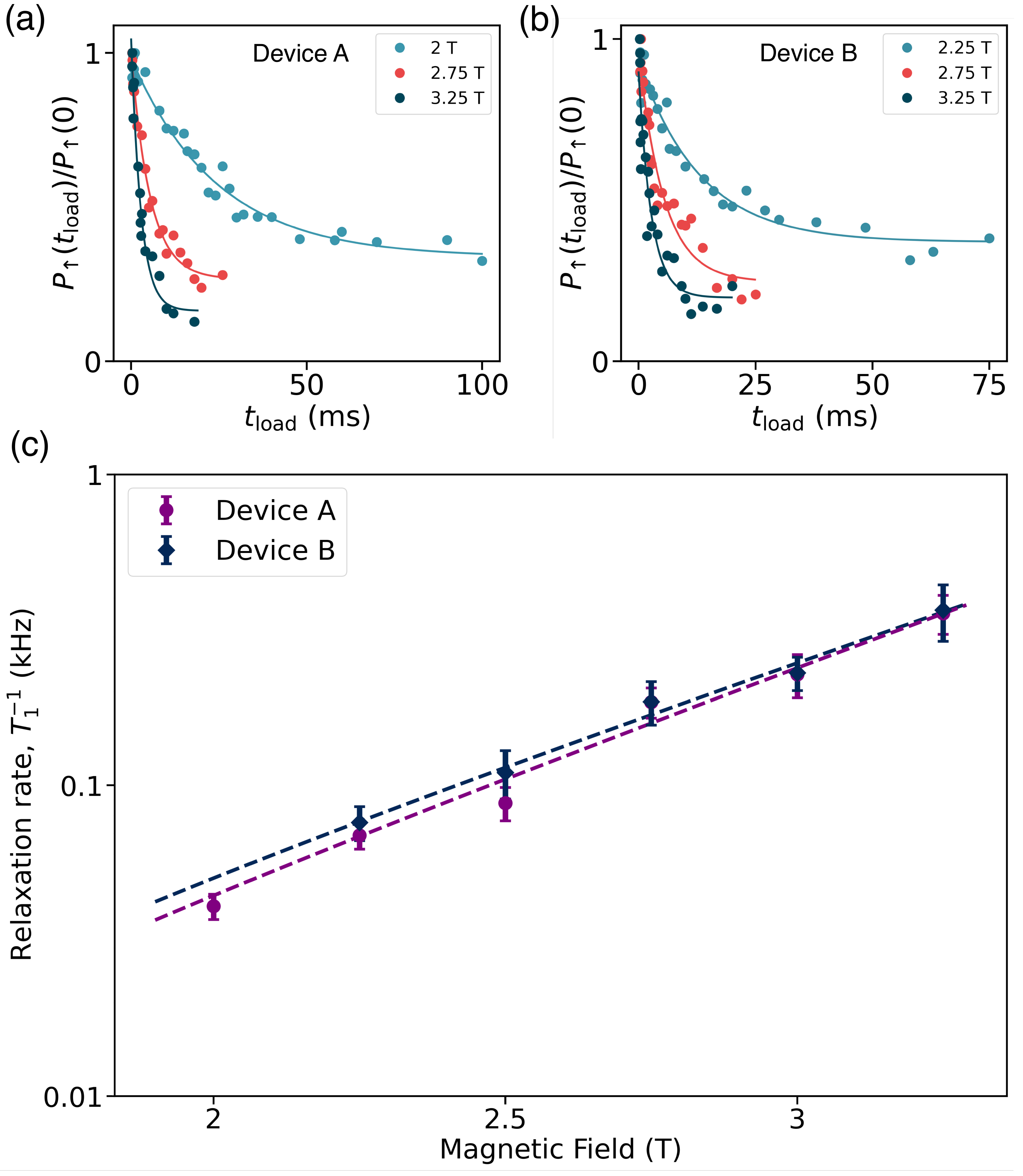}
    \caption{\textbf{Spin relaxation rate.} Normalized $\ket{\uparrow}$ state fraction as a function of $t_{\text{load}}$ at three magnetic fields for device A (a) and device B (b). Each plot it fitted to an exponential decay to obtain time constant $T_1$.
    Residual $\ket{\uparrow}$ fraction could be due thermal spin revivals or long lifetime $\ket{\uparrow}$ states, both of which are more likely at low field.
    c) Relaxation rate dependence on magnetic field for both devices plotted with a logarithmic y-axis.}
    \label{fig:relax rate}
\end{figure}

We next characterize the spin relaxation time, $T_1$, by measuring the probability of observing a $\ket{\uparrow}$ electron in the `read' phase as a function of `load’ duration, $t_\text{load}$. We observe an exponential decay of the $\ket{\uparrow}$ electron probability as a function of time in Fig. \ref{fig:relax rate}a and b, for device A and B respectively, as the $\ket{\uparrow}$ electrons becomes more likely to relax to the $\ket{\downarrow}$ ground state at longer load periods. To extract the $\ket{\uparrow}$ electron probability, we use a modified detection method that labels the spin $\ket{\uparrow}$ tunneling events relative to the spin $\ket{\downarrow}$ event (see Methods). This approach, that labels spins relative to a reference point unique to each trace, makes the method more robust to low frequency detuning fluctuations of the QD than the static time threshold discussed above. This feature becomes particularly relevant for long measurement ($>20$~ms) where charge jumps are more likely to occur, shifting the optimal time threshold over the course of the experiment.

%We calculate the $\ket{\uparrow}$ electron probability using the Classifier method to mitigate effects of charge noise which is particularly detrimental for long $t_{\text{load}}$ measurements. %Add: Explain residual spin-ups? Why have a residual of 5\% spin-ups? 

We repeat the measurement at different magnetic fields and plot its impact on relaxation rate $T_1^{-1}$, see  Fig.~\ref{fig:relax rate}c. We observe a decrease in the relaxation rate as the field is decreased, resulting in a maximum measured relaxation time of $T_1 = 24.5 \pm 2.3$~ms at $B = 2$~T for device A and $T_1 = 13.2 \pm 1.7$~ms at $B = 2.25$~T for device B.  We use the functional dependence of the spin relaxation rate on magnetic field to distinguish the dominant relaxation mechanisms.
%We find 
The relationship is well described by the expression $T_1^{\ -1} = K_{\text{J}} B^3 + K_{\text{ph}} B^7$ that includes spin-orbit mediated relaxation such as the effect of Johnson noise, with a $B^3$ dependence \cite{Huang2014}, and phonon-induced relaxation, scaling with $B^7$ \cite{Tahan2014}. We provide the coefficients for the fit in Table \ref{tab:T1K}. The results suggest that Johnson noise is dominant at the low end of the field range whereas phonon noise dominates at the high field end. %, expected for low fields around 2~T in which the Zeeman energy is lower than the predicted valley splitting \cite{Huang2014}. 
No evidence of relaxation hotspots suggests a valley splitting either lower than 230~$\mu$eV or higher than 350~$\mu$eV \cite{Yang2013}. We hypothesize that the similar $T_1^{\ -1}$ rates for both devices, which is determined by the symmetry of the spin-orbit coupling, indicates the QDs may have similar shapes and hence form in similar locations of the device~\cite{Virginia2021}.
%Effect of 1/f noise or magnetic noise are not conisdered due to their longer time scales/small effect. 
The value of the relaxation time at $B =$ 2~T in device A is comparable to other silicon MOS devices \cite{Petit2018,Virginia2021} and significantly larger than the typical dephasing times $T_2^*$ in silicon QDs \cite{Veldhorst2014, steinacker2024, Irene2025}. Identical relaxation rates could facilitate simultaneous state preparation of multiple devices using the same voltage pulse timings.

\begin{table}[!ht]
    \centering
    \renewcommand{\arraystretch}{1.4}
    \setlength{\tabcolsep}{10pt}
    \begin{tabular}{ccc}
        \hline\hline
        Device & $K_{\text{J}}$ (Hz/T$^3$) & $K_{\text{ph}}$ (Hz/T$^7$) \\
        \hline
        A & $4.7 \pm 0.9$ & $0.05 \pm 0.01$ \\
        B & $5.6 \pm 0.9$ & $0.04 \pm 0.01$ \\
        \hline\hline
    \end{tabular}
    \caption{Fitted parameters to the relaxation rate magnetic field dependence.}
    \label{tab:T1K}
\end{table}

\section{Readout Visibility}\label{sec:fidelity}

Having established the spin-dependent nature of the tunneling events, we move on to discussing the readout visibility. We use an analytical model developed in Ref.~\cite{Keith2022} that takes into account the time dependence of the tunnel rates due to the ramp in the `read' phase. The distribution of tunneling out events for the spin $\ket{\uparrow}$ and spin $\ket{\downarrow}$, is given by the probability density function (PDF) of the tunneling process (non-homogeneous Poisson process):   

\begin{equation}
  f_{\uparrow / \downarrow}(t) =
\frac{\Gamma  {e}^{x^{\uparrow/\downarrow} } }
{{e}^{x^{\uparrow/\downarrow} } + 1}
\left[ 
\frac{{e}^{\left(x^{\uparrow/\downarrow}  - \frac{rt}{k_\text{B} T_\text{e}}\right) } + 1}
{{e}^{x^{\uparrow/\downarrow} } + 1} 
\right]^{\frac{\Gamma k_\text{B} T_\text{e}}{r}}.
\label{eq:f^up/down}
\end{equation}

%\begin{equation}
%  f_{\uparrow / \downarrow}(t) =
%\frac{\Gamma }
%{{e}^{-x} + 1}
%\left( 
%\frac{{e}^{x} + 1}
%{{e}^{x - \frac{rt}{k_\text{B} T_\text{e}}} + 1} 
%\right)^{- \frac{\Gamma k_\text{B} T_\text{e}}{r}},
%\label{eq:f^up/down}
%\end{equation}

Here, $\Gamma$ is the maximum tunnel rate of the electron at large positive detuning and 

\begin{equation}
  x^{\uparrow/\downarrow} = \frac{\epsilon_0^{\uparrow/\downarrow} + rt}{k_B T_e},  
\end{equation}

\noindent denotes the normalised detuning relative to the Fermi-level at time $t$. The detuning is ramped at a rate $r = \epsilon_\text{ramp}/t_\text{ramp}$, where $\epsilon_\text{ramp}$ and $t_\text{ramp}$ are the ramp amplitude and time (see Methods). The spin-dependent energy levels are related by $\epsilon_0^{\uparrow} = \epsilon_0^{\downarrow} + E_\text{Z}$ through the Zeeman energy, and $T_\text{e}$ denotes the electron temperature of the reservoir.

%\noindent for the spin $\ket{\uparrow}$ and spin $\ket{\downarrow}$, respectively. Here, $\Gamma$ is the maximum tunnel rate of the electron at large positive detuning and $x$ is the electron detuning position at time $t$ relative to the Fermi-level, ramped from an initial detuning $\epsilon_0^{\uparrow/\downarrow}$ at a rate of $r = \epsilon_\text{ramp}/t_\text{ramp}$, where $\epsilon_\text{ramp}$ and $t_\text{ramp}$ are the ramp amplitude and time. Here, $T_\text{e}$ is the temperature of the Fermi reservoir. 

To test whether the model describes our data, we fit a generalized PDF including the $\ket{\uparrow}$ and $\ket{\downarrow}$ state probability outcomes to the histograms in Fig.~\ref{fig:peak sep}a,b:

\begin{equation}
f_{\text{total}}(t) = \frac{N_{\uparrow}}{N_{\text{total}}}f_{\uparrow}(t) 
+ \frac{(1-N_\uparrow)}{N_{\text{total}}} f_{\downarrow}(t).
% f^{total}(t) = \frac{N_{\uparrow}}{N_{total}} \left( f^{\uparrow}(t) + \exp\left( \frac{-t_m}{T_1} \right) f^{\uparrow}(t) \right) \\
% + \frac{N_{\uparrow}}{N_{total}} f^{\downarrow}(t).
\label{eq:f^total}
\end{equation}

\begin{figure}[t!]
    \includegraphics[width=\columnwidth]{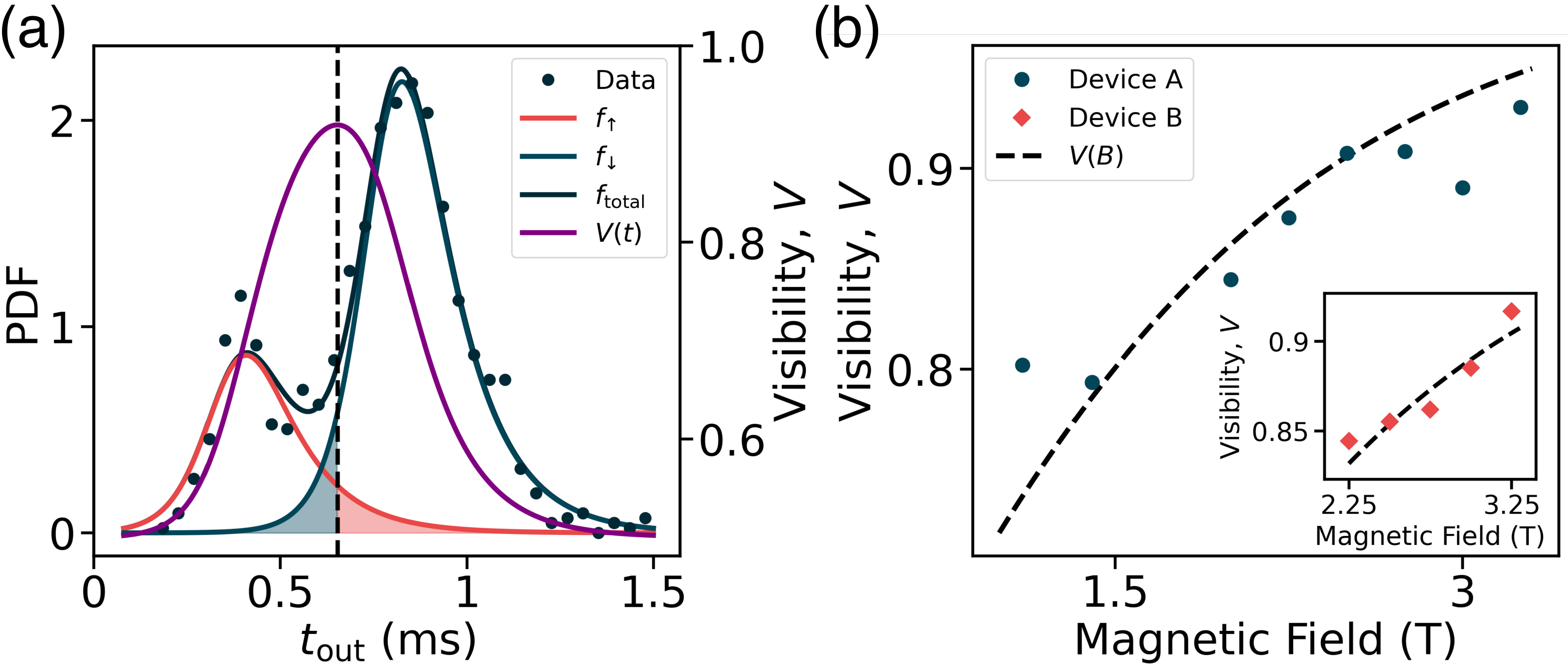}
    \caption{\textbf{Spin Readout Fidelity.} a) Histogram of $t_{\text{out}}$ in Fig.~\ref{fig:peak sep}b fit to the analytical model. Dashed line indicates a maximum visibility of 92\% at optimal time threshold $t = 0.65$~ms and the shaded regions show the areas of error. b) Visibility as a function of magnetic field for both devices obtained from the fits to the data in Fig.~\ref{fig:peak sep}.}
    \label{fig:fidelity}
\end{figure}

The good quality of the fit confirms the validity of the model (see Fig.~\ref{fig:peak sep}a,b and Fig.~\ref{fig:fidelity}a). We use the extracted fit parameters to determine the readout fidelity of each spin state, which can be understood as the proportion of correctly labeled spin states given a time threshold. Therefore, it can be written in terms of the normalized cumulative density function, %using $f^{\uparrow / \downarrow}(t)$, 

\begin{equation}
    C_{\uparrow / \downarrow}(t) = 1 - \left[ \frac{{e}^{\left(x^{\uparrow/\downarrow} - \frac{rt}{k_\text{B} T_\text{e}}\right)} + 1}
    {{e}^{x^{\uparrow/\downarrow}} + 1}\right]^{\frac{\Gamma k_B T_e}{r}},
\label{eq:CDF}
\end{equation}

and we use the visibility metric defined as~\cite{Barthel2009},

\begin{equation}
V(t) = \frac{1}{2}\left[C_{\uparrow}(t) - C_{\downarrow}(t)+1\right].
\label{eq:visible}
\end{equation}

We carry out the analysis for each magnetic field using
%an optimal time threshold determined analytically for the point at which the vsibility reaches maximum value, indicated.. 
a time threshold defined as the point at which the visibility reaches its maximum value, as shown in Fig. \ref{fig:fidelity}a. We calculate a visibility of 93.1\% for device A and 91.7\% for device B at $B = 3.25$~T. The reduced visibility at lower magnetic fields is an effect predicted by the RSM \cite{Keith2022} and is observed in both devices in Fig.~\ref{fig:fidelity}b. We determine the visibility at the optimal time threshold for a range of magnetic fields, $V(B)$, using an average of the fit parameters and plot in Fig. \ref{fig:fidelity}b. This predicts $V > 98$~\% at $B = 4.5$~T for device A and $B = 5.9$~T for device B (considering no $T_1$ errors). Besides increasing the magnetic field, higher readout visibility could be obtained by reducing the electron temperature or increasing the electron tunnel rates.

%Fidelity of the classifier method is assessed by building a numerical model using a time-varying transition matrix. Method is applied to 10,000 simulated traces to determine the probability that a spin state is correctly labeled, $F_{\uparrow/\downarrow}$. Simulation parameters are obtained from the analytical fits and validated by applying the threshold method to reproduce the experimental distribution in Fig. 2a, as shown in Fig. \ref{fig:fidelity}a. TODO: datapoint at B=1.4T for classifier.

\section{\label{sec:5}Conclusion}

We have performed the first demonstration of electron spin readout in an integrated circuit, manufactured using an industry-standard process and containing classical addressing electronics, i.e. a multiplexer. Using the MUX, we tested three nominally identical devices, two of which showed the spin readout results presented in this Article. The third one showed single-shot charge readout but no spin-dependent transitions which we attribute to the higher charge noise on that cell. Further data analysis methods improving on the dynamical thresholding may be necessary to reveal spin dependence. 

Moving forward, these results encourage the incorporation of isotopically enriched silicon-on-insulator in the fabrication process to enable the demonstration of high-fidelity single-qubit gates using the very same unit cells. In addition, we foresee that more complex unit cells may be explored including double QD structures, to enable spin-dependent readout via Pauli spin blockade and the demonstration of two-qubit interactions. The use of the MUX should allow rapid exploration of the parameter space to identify the critical device dimensions. Finally, we anticipate that the recent demonstration of state-of-the-art charge sensitivity achieved with an integrated superinductor~\cite{swift2025} may enable the integration not only of the classical addressing electronics but also the readout peripherals all on the same chip alongside the spin readout unit. Our results open a path to reduce system complexity by integrating more elements on chip enabling, as much as possible, a transition from modular quantum computing systems to integrated semiconductor-based quantum computers. 

%\begin{itemize}
%    \item $^{28}$Si
%    \item Combined demonstration with TiN superinductor for a fully integrated demonstration including the readout and multiplexing electronics as well as the qubit unit cell. 
%    \item More complex unit cells with a sensor an double quantum dot to demonstrate Pauli spin blockade and two-qubit interactions.
%\end{itemize}

%\section{Acknowledgements}

%\section{Author Contributions}

%\section{Competing interests}

\section{\label{sec:Methods} Methods}

\subsection{\label{append:Fab} Device fabrication} 
The devices were fabricated using the GlobalFoundries 22~nm FDSOI process (22FDX). All the devices in this study were contained in a single die.

\subsection{\label{append:Setup}Measurement set-up} 
Measurements are performed in a BlueFors LD dilution refrigerator at a base temperature of $T\sim 10$\,mK. The IC is glued to a printed circuit board (PCB) and connected to the off-chip LC resonant circuit by aluminum bond wires. A QDevil QDAC 1 is used to supply dc voltages to the device terminals and chip power supplies. Fast voltage pulses are applied to the device gates through attenuated coaxial lines with an Agilent 33522A Arbitrary Waveform Generator (AWG). The rf signal for rf-reflectometry readout is generated by a Rohde \& Schwarz SMC100A signal generator and transmitted through attenuated and filtered coaxial lines to the chip. The dc and rf signals are combined through bias-Ts on the PCB. The reflected rf signal is amplified at 4\,K and room temperature, and magnitude and phase are obtained using quadrature demodulation. Signals are further amplified and filtered by Stanford SR560 preamplifiers then digitised via a Spectrum M4i digitiser card. An external magnetic field is applied in the $z-$axis of an AMI430 magnet. 

% We perform the measurements at the base temperature of a dilution refrigerator ($T\sim 10$\,mK). We send low-frequency signals through cryogenic low-pass filters with a cut-off frequency of 65~kHz, while we apply pulsed signals through attenuated coaxial lines. Both signals are combined through bias-Ts at the sample PCB (printed circuit-board) level. The PCB was made from RO4003C 0.8\,mm thick with an immersion silver finish. For readout, we use radio-frequency reflectometry applied on the ohmic contact of the device. We send radio-frequency signals through attenuated coaxial lines to an on-PCB $LC$ resonator, arranged in a parallel configuration, formed by a coupling capacitor ($C_\text{c}$), a 100~nm-thick NbTi superconducting spiral inductor ($L$) and the parasitic capacitance to ground ($C_\text{p}$), as shown in Fig.~\ref{Fig:1}(c). We drive the resonator at 512.25~MHz which is the frequency of the system when G$_\text{S}$ is well above threshold. The reflected rf signal is then amplified at 4\,K and room temperature, followed by quadrature demodulation, from which the amplitude and phase of the reflected signal were obtained (homodyne detection). 

\subsection{\label{append:MUX}Multiplexer design} 

To access the devices in this study, we use an on-chip one-to-128 multiplexer with a 7-bit digital select bus and 6-wire analog signal bus  that enables the sequential selection of a single cell among 128 different device variants. The device source, drain, sensor gate and qubit gate are connected to the 6-wire bus. The multiplexed element is implemented using complementary transmission gates --  back-to-back N-type and P-type field-effect transistors -- with forward back-bias capabilities, followed by a pull-down switch on the device side, which connects the terminals of the non-selected devices to a global ground (VSS$_\text{G}$). The n-type back-gate is common to all devices. For additional details, see Ref.~\cite{Thomas2025}.

\subsection{\label{appen:qd parameters}Quantum dot parameters}

\subsubsection{Reservoir temperature}
\begin{figure}[ht]
    \includegraphics[width=\columnwidth]{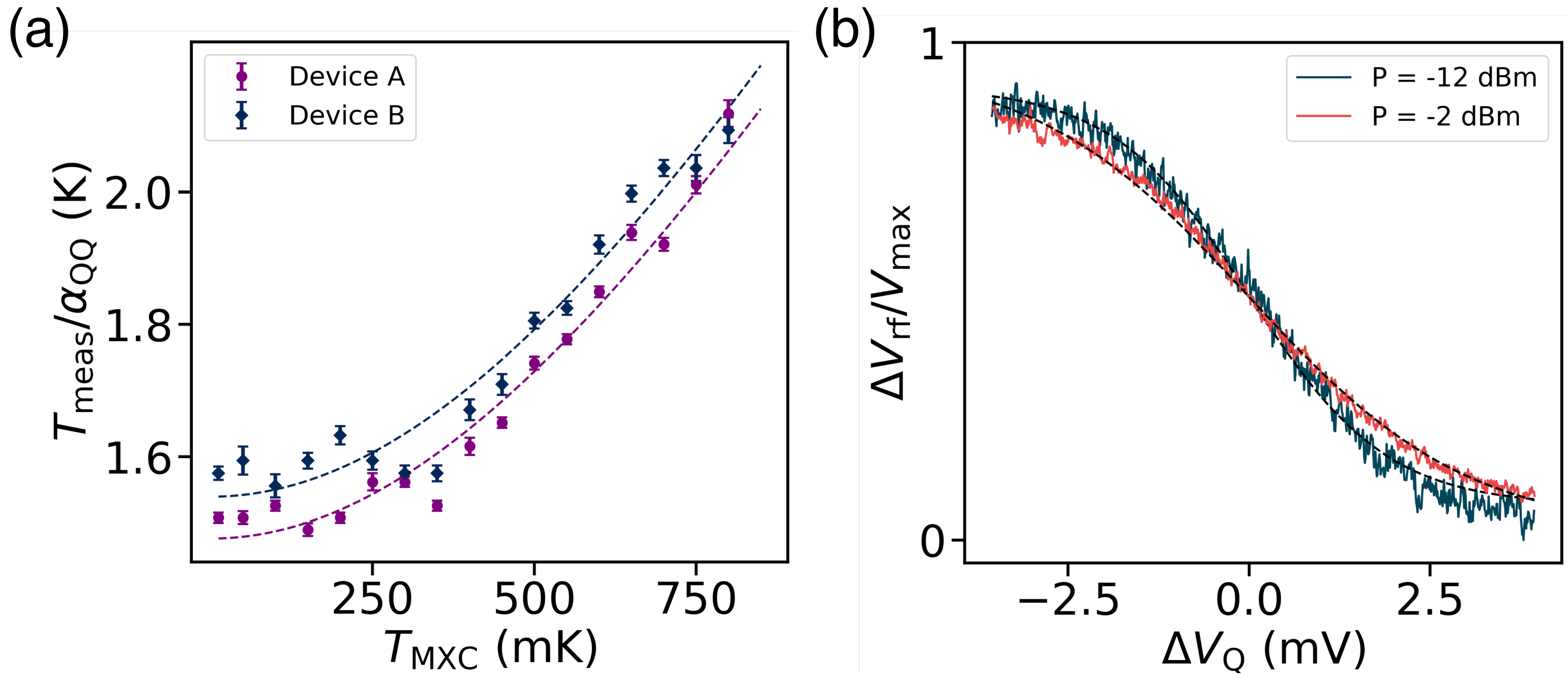}
    \caption{\textbf{Reservoir temperature and qubit dot lever arm.} a) Width of the thermal broadening for the qubit dot unloading signal as a function of fridge temperature. %Reservoir temperature and qubit dot lever arm to gate $G_\text{Q}$ are within the error of each device. 
    b) Qubit dot unloading signal during the diagonal ramped `read' stage at rf power -2 dBm and -12 dBm. %$\Delta V_\text{Q}$ is obtained using $\Delta t \times  \epsilon_\text{ramp}/(t_\text{ramp}/k_\text{B}$.
    }
    \label{fig:alpha thermo}
\end{figure}

We determine the effective reservoir temperature of the device using a thermometry measurement, see Fig.~\ref{fig:alpha thermo}a. A vertical ramp across the charge transition is implemented using the qubit gate $G_\text{Q}$ and the qubit dot occupation signal is fitted to a Fermi-Dirac distribution, see Fig.~\ref{fig:alpha thermo}b. We relate the width of the distribution to the measured reservoir temperature, $T_\text{meas}$, using $\Delta V_\text{Q}$=$k_\text{B} T_{\text{meas}}/e\alpha_\text{QQ}$ where $\alpha_\text{QQ}$ is the qubit dot lever arm to qubit gate $G_\text{Q}$ which apriori is unknown. To determine $T_\text{meas}$ and $\alpha_\text{QQ}$, we increase the mixing chamber temperature, $T_{\text{MXC}}$ and monitor the Fermi width. We observe the width is constant at low fridge temperatures, due to the effective base electron temperature of the reservoir, $T_\text{eff}$, to then increase linearly with a gradient determined by $\alpha_\text{QQ}$. We then fit the data in Fig.~\ref{fig:alpha thermo}a to equation,

\begin{equation}
    \frac{T_\text{meas}}{\alpha_\text{QQ}}=\frac{1}{\alpha_\text{QQ}} \sqrt{T_{\text{eff}}^2 + T_{\text{MXC}}^2},
\label{eq:thermo}
\end{equation}

\noindent and obtain $T_{\text{eff}}$ = 821 $\pm$ 28~mK for device A and $T_{\text{eff}}$ = 840 $\pm$ 35~mK for device B, respectively. We discuss $\alpha_\text{QQ}$ in the next section. 
%Values are within error of each other as expected. 
We determine that the high effective base electron temperature of the reservoir is due to power broadening from the rf-reflectometry signal. In Fig.~\ref{fig:alpha thermo}b we show that increasing the applied rf power of the rf generator, $P$, increases the broadening of the qubit dot occupation signal during a diagonal `read' ramp. Spin readout experiments and reservoir temperature characterization are performed at a high rf power of $P =$ -2~dBm.  

\subsubsection{Gate lever arms}
We measure the gate lever arms of the sensor and qubit dots separately.
%The lever arm converts the voltage applied to each gate to the electrostatic energy at the dot.
The lever arm of the sensor dot to the gate $G_\text{S}$, $\alpha_\text{SS}$, is obtained using an SET Coulomb diamond with $V_\text{Q}$ set close to the charge transition chosen for spin readout. The qubit dot lever arm to the gate $G_\text{Q}$, $\alpha_\text{QQ}$, is extracted from the fit in the thermometry measurement in Fig.~\ref{fig:alpha thermo}a. The cross-coupled lever arms, $\alpha_\text{QS}$ and $\alpha_\text{SQ}$ are obtained using the dot-to-reservoir gradients extracted from the map in Fig.~\ref{fig:device setup}b. The lever arms for both devices are given in Table~\ref{tab:lever_arms}.

% Device A has the final lever arm matrix 

% \begin{equation}
%     \begin{pmatrix}
%     \alpha_\text{QQ} & \alpha_\text{QS} \\
%     \alpha_\text{SQ} & \alpha_\text{SS} 
%     \end{pmatrix}
%     =
%     \begin{pmatrix}
%     0.56 \pm 0.02 & 0.047 \pm 0.001 \\
%     0.013 & 0.46 
%     \end{pmatrix}
% \label{eq:mu matrix}
% \end{equation}

% and using equivalent methods for device B results in lever arm matrix

% \begin{equation}
%     \begin{pmatrix}
%     \alpha_\text{QQ} & \alpha_\text{QS} \\
%     \alpha_\text{SQ} & \alpha_\text{SS} 
%     \end{pmatrix}
%     =
%     \begin{pmatrix}
%     0.55 \pm 0.02 & 0.046 \pm 0.002 \\
%     0.05 & 0.75 
%     \end{pmatrix}
% \label{eq:mu matrix}
% \end{equation}

\begin{table}[h]
    \centering
    \renewcommand{\arraystretch}{1.4}
    \setlength{\tabcolsep}{10pt} % spacing between columns
    \begin{tabular}{lcc}
        \hline\hline
        & Device A & Device B \\
        \hline
        $\alpha_\text{QQ}$ & $0.56 \pm 0.02$ & $0.55 \pm 0.02$ \\
        $\alpha_\text{QS}$ & $0.047 \pm 0.001$ & $0.046 \pm 0.002$ \\
        $\alpha_\text{SQ}$ & $0.013$ & $0.05$ \\
        $\alpha_\text{SS}$ & $0.46$ & $0.75$ \\
        \hline\hline
    \end{tabular}
    \caption{Lever arms for both devices.}
    \label{tab:lever_arms}
\end{table}

\subsubsection{Effective ramp rate}
The voltage vector 

\begin{equation}
    \begin{pmatrix}
    V_\text{Q} \\
    V_\text{S}
    \end{pmatrix}
    = 
    \begin{pmatrix}
    - 2.05  \\ %\pm 0.01
    0.505 %\pm 0.001
    \end{pmatrix} 
    \ \text{mV},
\label{eq:mu matrix}
\end{equation}

is applied to the device gates to ramp diagonally across the charge transition. 
%A conversion factor between the voltage applied the AWG and the amplitude arriving at the device through the attenuated pulse line is applied.
%determined by applying a square wave pulse to $G_\text{Q/S}$ and measuring the splitting of an SET Coulomb peak, shown in Fig. \ref{fig:pulse line}. 
The effective detuning ramp experienced by each dot can be calculated using the formula:

% \begin{figure}[h!]
%     \includegraphics[width=\columnwidth]{Figures/Appendix pulse lines.png}
%     \caption{\textbf{Pulse line characterization.} a) AWG voltage of 1.74~V splits an SET Coulomb peak by 8.91 $\pm$ 0.05~mV when applied to the $G_\text{Q}$ pulse line. b) AWG voltage of 0.12~V splits an SET Coulomb peak by 3.69 $\pm$ 0.01~mV when applied to the $G_\text{S}$ pulse line. Amplitude at device obtained by fitting a Gaussian to each peak.}
%     \label{fig:pulse line}
% \end{figure}

\begin{equation}
    \begin{pmatrix}
    \epsilon_\text{Q} \\
    \epsilon_\text{S}
    \end{pmatrix}
    = \begin{pmatrix}
    \alpha_\text{QQ} & \alpha_\text{QS} \\
    \alpha_\text{SQ} & \alpha_\text{SS} 
    \end{pmatrix}
    \begin{pmatrix}
    V_\text{Q} \\
    V_\text{S}
    \end{pmatrix}.
\label{eq:mu matrix2}
\end{equation}

The diagonal detuning ramp, $\epsilon_\text{ramp}$, is equivalent to the detuning of the qubit dot, $\epsilon_\text{Q}$, relative to the reservoir, $\epsilon_{\text{res}}$. Setting $\epsilon_{\text{res}} =$ 0 eV, the ramp applies a detuning of $|\epsilon_\text{ramp}| =$ 1.12 $\pm$ 0.03~meV. 
Converting from time into energy during the `read' stage is achieved using $\Delta E = \Delta t \times  \epsilon_\text{ramp}/t_\text{ramp}$. Equivalent methods are used for device B.

\subsection{\label{appen:Final exit referenced classifier}Final exit referenced classifier}

In long experiments such as spin relaxation measurements, slow charge noise and detuning drifts  become particularly pronounced. To mitigate these effects, we classify spin states relative to the electron’s final exit from the dot. The algorithm first determines the time at which the electron leaves the dot permanently, $t_{\downarrow\text{out}}$, using a penalized change-point detection~\cite{burg2022evaluationchangepointdetection}. The signal preceding this point resembles a standard ERO-like measurement trace. If the signal `blips’ above a voltage threshold, the shot is classified as a $\ket{\uparrow}$ electron and time $t_{\uparrow \text{out}}$ is obtained. To prevent misclassifications of thermally activated back-tunneling events from spin $\ket{\downarrow}$ states, any blip occurring within a $1.75\,k_\mathrm{B}T_\mathrm{eff}$ energy window of $t_{\downarrow\text{out}}$ is disregarded. By referencing each trace to its own final exit time, makes this classifier resilient to slow detuning drifts that modify the effective detuning $\varepsilon_{\uparrow/\downarrow}$ levels.

\subsection{\label{appen:initialization}Spin initialization}

\begin{figure}[t!]
    \includegraphics[width=\columnwidth]{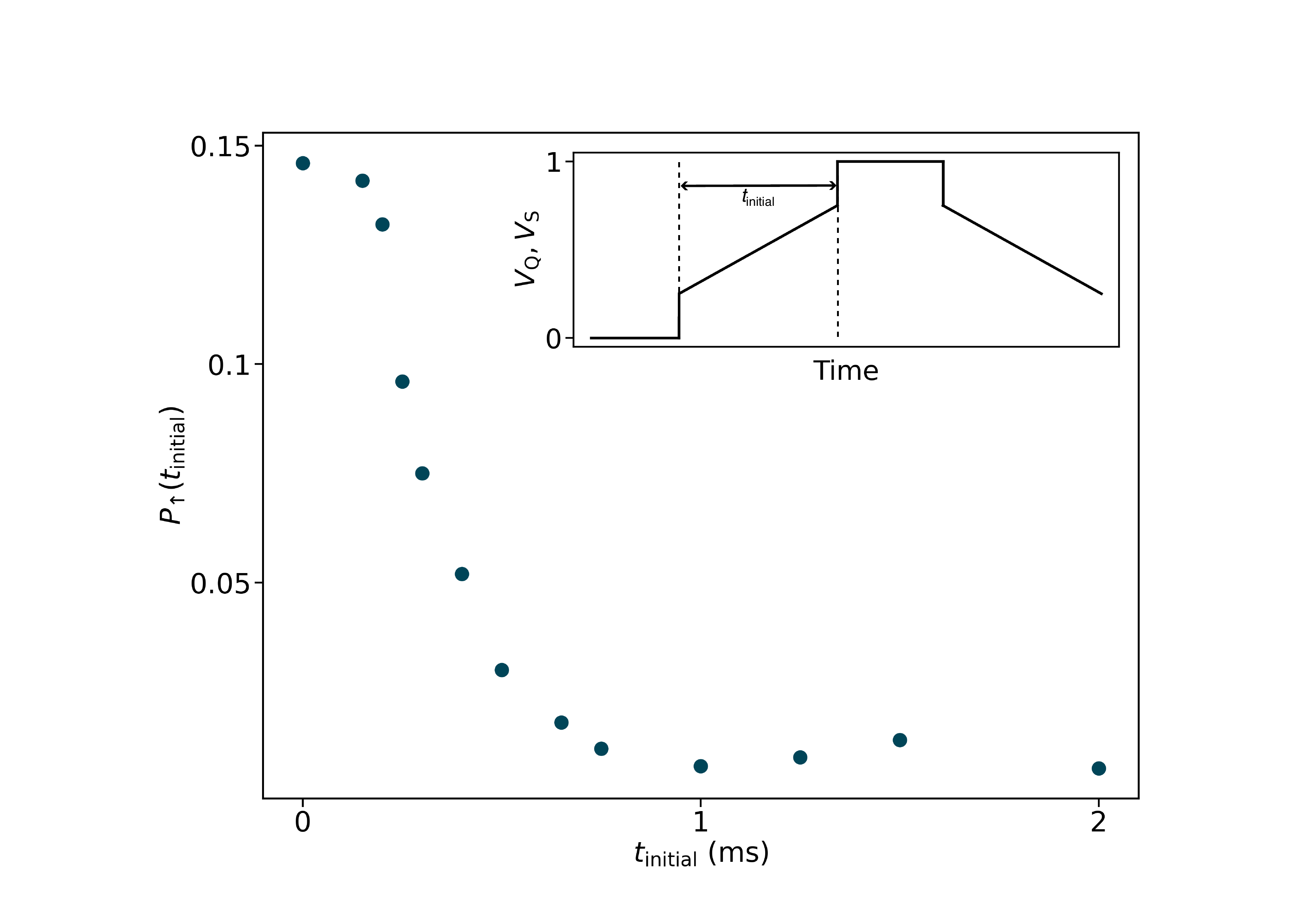}
    \caption{\textbf{Spin initialization.} The $\ket{\uparrow}$ state fraction as a function of initial ramp time $t_{\text{initial}}$. Ramp amplitude remains constant for each data point. Inset shows the voltage pulse sequence applied to gates $G_\text{Q}$ and $G_\text{S}$.}
    \label{fig:initial}
\end{figure}

We perform an initialization ramp in the opposite direction to the `read' stage to deterministically load an electron in the $\ket{\downarrow}$ state. Fig. \ref{fig:initial} shows the $\ket{\uparrow}$ fraction 
% prob of observing a spin-up electron
as a function of initial ramp time, $t_{\text{initial}}$. As the ramp time increases compared to the electron tunnel rates, there is a higher probability of the electron loading onto the lower $\ket{\downarrow}$  electrochemical potential level during the ramp. Data is obtained at $B =$ 3~T and the $\ket{\uparrow}$ fraction evaluated using the threshold method with a time domain threshold set at minimum overlap of the distribution peaks. 
The ability to prepare a $\ket{\downarrow}$  state with > 95\% certainty holds promise for future spin control experiments.

\section*{Data Availability}
\noindent The data supporting the plots within this article and other findings of this study are available from the corresponding authors upon reasonable request.

\section*{Acknowledgements}
\noindent We thank Nigel Cave at GlobalFoundries for helpful discussions. We acknowledge technical support and helpful conversations with James Kirkman and Mathieu de Kruijf at Quantum Motion and technical support from G.~Antilen Jacob at the London Centre for Nanotechnology.  I.C.C.~acknowledges support from the Engineering and Physical Sciences Research Council (EPSRC) through the Hub for Qantum Computing and Simulations (HQCS). A.G.-S. acknowledges an Industrial Fellowship from the Royal Commission for the Exhibition of 1851. M.F.G-Z.~acknowledges support from the UKRI Future Leaders Fellowship (MR/V023284/1).

\section*{Author contributions}
\noindent I.C.C. conducted the experiments and analysed the results presented in this work with input from V.C-T. G.M.N., A.G-S, J.J.L.M. and M.F.G-Z. V.C-T. conducted the single-shot data analysis. D.J.I., T.H.S and M.A.I.J conducted preliminary experiments. A.G-S and M.F.G-Z conceived the quantum dot devices. A.G-S designed the IC with input from R.C.C.L. and M.F.G-Z. J.J.L.M. and M.F.G-Z. supervised the work. All authors contributed to the writing of the manuscript. 

%J.F.C.W., M.A.F. and R.C.C.L. conducted preliminary experiments; T.M. and G.A.O developed the quantum dot rf simulator in Supplementary Section 4 informed by electrostatic modelling from J.W.; T. M. performed and wrote about the simulations under the supervision of D. F. W. and M. F. G. Z.; F.E.v.H. characterised the superconducting resonator{\color{blue} , and carried out experiments on further devices to validate the rf-cascade technique with support from J.F.C.W and M.F.G.Z.;} S.M.P. designed the device under the supervision of M.A.F., M.F.G.Z. and J.J.L.M.; J.J. and S.K. fabricated the device under the supervision of B.G.; N.J. maintained the experimental setup; J.F.C.W., M.F.G.Z. and J.J.L.M. wrote the manuscript with input from R.C.C.L., M.A.F., N.J. and S.M.P.; M.F.G.Z. and J.J.L.M. conceived and oversaw the experiment.

\section*{Competing Interests}
\noindent A.G-S. and M.F.G-Z are inventors on patent application [EP25382976] which describes the physical implementation of the quantum dot device used in this work.

% \newpage
% \bibliography{General} 

\begin{thebibliography}{44}%
\makeatletter
\providecommand \@ifxundefined [1]{%
 \@ifx{#1\undefined}
}%
\providecommand \@ifnum [1]{%
 \ifnum #1\expandafter \@firstoftwo
 \else \expandafter \@secondoftwo
 \fi
}%
\providecommand \@ifx [1]{%
 \ifx #1\expandafter \@firstoftwo
 \else \expandafter \@secondoftwo
 \fi
}%
\providecommand \natexlab [1]{#1}%
\providecommand \enquote  [1]{``#1''}%
\providecommand \bibnamefont  [1]{#1}%
\providecommand \bibfnamefont [1]{#1}%
\providecommand \citenamefont [1]{#1}%
\providecommand \@href[1]{\@@startlink{#1}\@@href}%
\providecommand \@@href[1]{\endgroup#1\@@endlink}%
\providecommand \@sanitize@url [0]{\catcode `\\12\catcode `\$12\catcode `\&12\catcode `\#12\catcode `\^12\catcode `\_12\catcode `\%12\relax}%
\providecommand \@@startlink[1]{}%
\providecommand \@@endlink[0]{}%
\providecommand \@url [1]{\endgroup\@href {#1}{\urlprefix }}%
\providecommand \urlprefix  [0]{URL }%% 
\providecommand \doibase [0]{https://doi.org/}%
\providecommand \selectlanguage [0]{\@gobble}%
\providecommand \bibinfo  [0]{\@secondoftwo}%
\providecommand \bibfield  [0]{\@secondoftwo}%
\providecommand \translation [1]{[#1]}%
\providecommand \BibitemOpen [0]{}%
\providecommand \bibitemStop [0]{}%
\providecommand \bibitemNoStop [0]{.\EOS\space}%
\providecommand \EOS [0]{\spacefactor3000\relax}%
\providecommand \BibitemShut  [1]{\csname bibitem#1\endcsname}%
\let\auto@bib@innerbib\@empty
%</preamble>
\bibitem [{\citenamefont {Madsen}\ \emph {et~al.}(2022)\citenamefont {Madsen}, \citenamefont {Laudenbach}, \citenamefont {Askarani}, \citenamefont {Rortais}, \citenamefont {Vincent}, \citenamefont {Bulmer}, \citenamefont {Miatto}, \citenamefont {Neuhaus}, \citenamefont {Helt}, \citenamefont {Collins}, \citenamefont {Lita}, \citenamefont {Gerrits}, \citenamefont {Nam}, \citenamefont {Vaidya}, \citenamefont {Menotti}, \citenamefont {Dhand}, \citenamefont {Vernon}, \citenamefont {Quesada},\ and\ \citenamefont {Lavoie}}]{Madsen2022}%
  \BibitemOpen
  \bibfield  {author} {\bibinfo {author} {\bibfnamefont {L.~S.}\ \bibnamefont {Madsen}}, \bibinfo {author} {\bibfnamefont {F.}~\bibnamefont {Laudenbach}}, \bibinfo {author} {\bibfnamefont {M.~F.}\ \bibnamefont {Askarani}}, \bibinfo {author} {\bibfnamefont {F.}~\bibnamefont {Rortais}}, \bibinfo {author} {\bibfnamefont {T.}~\bibnamefont {Vincent}}, \bibinfo {author} {\bibfnamefont {J.~F.~F.}\ \bibnamefont {Bulmer}}, \bibinfo {author} {\bibfnamefont {F.~M.}\ \bibnamefont {Miatto}}, \bibinfo {author} {\bibfnamefont {L.}~\bibnamefont {Neuhaus}}, \bibinfo {author} {\bibfnamefont {L.~G.}\ \bibnamefont {Helt}}, \bibinfo {author} {\bibfnamefont {M.~J.}\ \bibnamefont {Collins}}, \bibinfo {author} {\bibfnamefont {A.~E.}\ \bibnamefont {Lita}}, \bibinfo {author} {\bibfnamefont {T.}~\bibnamefont {Gerrits}}, \bibinfo {author} {\bibfnamefont {S.~W.}\ \bibnamefont {Nam}}, \bibinfo {author} {\bibfnamefont {V.~D.}\ \bibnamefont {Vaidya}}, \bibinfo {author} {\bibfnamefont {M.}~\bibnamefont {Menotti}}, \bibinfo {author}
  {\bibfnamefont {I.}~\bibnamefont {Dhand}}, \bibinfo {author} {\bibfnamefont {Z.}~\bibnamefont {Vernon}}, \bibinfo {author} {\bibfnamefont {N.}~\bibnamefont {Quesada}},\ and\ \bibinfo {author} {\bibfnamefont {J.}~\bibnamefont {Lavoie}},\ }\bibfield  {title} {\bibinfo {title} {Quantum computational advantage with a programmable photonic processor},\ }\href {https://doi.org/10.1038/s41586-022-04725-x} {\bibfield  {journal} {\bibinfo  {journal} {Nature}\ }\textbf {\bibinfo {volume} {606}},\ \bibinfo {pages} {75} (\bibinfo {year} {2022})}\BibitemShut {NoStop}%
\bibitem [{\citenamefont {Kim}\ \emph {et~al.}(2023)\citenamefont {Kim}, \citenamefont {Eddins}, \citenamefont {Anand}, \citenamefont {Wei}, \citenamefont {van~den Berg}, \citenamefont {Rosenblatt}, \citenamefont {Nayfeh}, \citenamefont {Wu}, \citenamefont {Zaletel}, \citenamefont {Temme},\ and\ \citenamefont {Kandala}}]{Kim2023}%
  \BibitemOpen
  \bibfield  {author} {\bibinfo {author} {\bibfnamefont {Y.}~\bibnamefont {Kim}}, \bibinfo {author} {\bibfnamefont {A.}~\bibnamefont {Eddins}}, \bibinfo {author} {\bibfnamefont {S.}~\bibnamefont {Anand}}, \bibinfo {author} {\bibfnamefont {K.~X.}\ \bibnamefont {Wei}}, \bibinfo {author} {\bibfnamefont {E.}~\bibnamefont {van~den Berg}}, \bibinfo {author} {\bibfnamefont {S.}~\bibnamefont {Rosenblatt}}, \bibinfo {author} {\bibfnamefont {H.}~\bibnamefont {Nayfeh}}, \bibinfo {author} {\bibfnamefont {Y.}~\bibnamefont {Wu}}, \bibinfo {author} {\bibfnamefont {M.}~\bibnamefont {Zaletel}}, \bibinfo {author} {\bibfnamefont {K.}~\bibnamefont {Temme}},\ and\ \bibinfo {author} {\bibfnamefont {A.}~\bibnamefont {Kandala}},\ }\bibfield  {title} {\bibinfo {title} {Evidence for the utility of quantum computing before fault tolerance},\ }\href {https://doi.org/10.1038/s41586-023-06096-3} {\bibfield  {journal} {\bibinfo  {journal} {Nature}\ }\textbf {\bibinfo {volume} {618}},\ \bibinfo {pages} {500} (\bibinfo {year}
  {2023})}\BibitemShut {NoStop}%
\bibitem [{\citenamefont {Bluvstein}\ \emph {et~al.}(2024)\citenamefont {Bluvstein}, \citenamefont {Evered}, \citenamefont {Geim}, \citenamefont {Li}, \citenamefont {Zhou}, \citenamefont {Manovitz}, \citenamefont {Ebadi}, \citenamefont {Cain}, \citenamefont {Kalinowski}, \citenamefont {Hangleiter}, \citenamefont {Bonilla~Ataides}, \citenamefont {Maskara}, \citenamefont {Cong}, \citenamefont {Gao}, \citenamefont {Sales~Rodriguez}, \citenamefont {Karolyshyn}, \citenamefont {Semeghini}, \citenamefont {Gullans}, \citenamefont {Greiner}, \citenamefont {Vuletic},\ and\ \citenamefont {Lukin}}]{Bluvstein2024}%
  \BibitemOpen
  \bibfield  {author} {\bibinfo {author} {\bibfnamefont {D.}~\bibnamefont {Bluvstein}}, \bibinfo {author} {\bibfnamefont {S.~J.}\ \bibnamefont {Evered}}, \bibinfo {author} {\bibfnamefont {A.~A.}\ \bibnamefont {Geim}}, \bibinfo {author} {\bibfnamefont {S.~H.}\ \bibnamefont {Li}}, \bibinfo {author} {\bibfnamefont {H.}~\bibnamefont {Zhou}}, \bibinfo {author} {\bibfnamefont {T.}~\bibnamefont {Manovitz}}, \bibinfo {author} {\bibfnamefont {S.}~\bibnamefont {Ebadi}}, \bibinfo {author} {\bibfnamefont {M.}~\bibnamefont {Cain}}, \bibinfo {author} {\bibfnamefont {M.}~\bibnamefont {Kalinowski}}, \bibinfo {author} {\bibfnamefont {D.}~\bibnamefont {Hangleiter}}, \bibinfo {author} {\bibfnamefont {J.~P.}\ \bibnamefont {Bonilla~Ataides}}, \bibinfo {author} {\bibfnamefont {N.}~\bibnamefont {Maskara}}, \bibinfo {author} {\bibfnamefont {I.}~\bibnamefont {Cong}}, \bibinfo {author} {\bibfnamefont {X.}~\bibnamefont {Gao}}, \bibinfo {author} {\bibfnamefont {P.}~\bibnamefont {Sales~Rodriguez}}, \bibinfo {author} {\bibfnamefont
  {T.}~\bibnamefont {Karolyshyn}}, \bibinfo {author} {\bibfnamefont {G.}~\bibnamefont {Semeghini}}, \bibinfo {author} {\bibfnamefont {M.~J.}\ \bibnamefont {Gullans}}, \bibinfo {author} {\bibfnamefont {M.}~\bibnamefont {Greiner}}, \bibinfo {author} {\bibfnamefont {V.}~\bibnamefont {Vuletic}},\ and\ \bibinfo {author} {\bibfnamefont {M.~D.}\ \bibnamefont {Lukin}},\ }\bibfield  {title} {\bibinfo {title} {Logical quantum processor based on reconfigurable atom arrays},\ }\href {https://doi.org/10.1038/s41586-023-06927-3} {\bibfield  {journal} {\bibinfo  {journal} {Nature}\ }\textbf {\bibinfo {volume} {626}},\ \bibinfo {pages} {58} (\bibinfo {year} {2024})}\BibitemShut {NoStop}%
\bibitem [{\citenamefont {Xu}\ \emph {et~al.}(2025)\citenamefont {Xu}, \citenamefont {Xia}, \citenamefont {Yu}, \citenamefont {Parakh}, \citenamefont {Khan}, \citenamefont {Megidish}, \citenamefont {You}, \citenamefont {Hemmerling}, \citenamefont {Jayich}, \citenamefont {Beck}, \citenamefont {Biener},\ and\ \citenamefont {H{\"a}ffner}}]{Xu2025}%
  \BibitemOpen
  \bibfield  {author} {\bibinfo {author} {\bibfnamefont {S.}~\bibnamefont {Xu}}, \bibinfo {author} {\bibfnamefont {X.}~\bibnamefont {Xia}}, \bibinfo {author} {\bibfnamefont {Q.}~\bibnamefont {Yu}}, \bibinfo {author} {\bibfnamefont {A.}~\bibnamefont {Parakh}}, \bibinfo {author} {\bibfnamefont {S.}~\bibnamefont {Khan}}, \bibinfo {author} {\bibfnamefont {E.}~\bibnamefont {Megidish}}, \bibinfo {author} {\bibfnamefont {B.}~\bibnamefont {You}}, \bibinfo {author} {\bibfnamefont {B.}~\bibnamefont {Hemmerling}}, \bibinfo {author} {\bibfnamefont {A.}~\bibnamefont {Jayich}}, \bibinfo {author} {\bibfnamefont {K.}~\bibnamefont {Beck}}, \bibinfo {author} {\bibfnamefont {J.}~\bibnamefont {Biener}},\ and\ \bibinfo {author} {\bibfnamefont {H.}~\bibnamefont {H{\"a}ffner}},\ }\bibfield  {title} {\bibinfo {title} {3{D}-printed micro ion trap technology for quantum information applications},\ }\href {https://doi.org/10.1038/s41586-025-09474-1} {\bibfield  {journal} {\bibinfo  {journal} {Nature}\ }\textbf {\bibinfo {volume}
  {645}},\ \bibinfo {pages} {362} (\bibinfo {year} {2025})}\BibitemShut {NoStop}%
\bibitem [{\citenamefont {Fowler}\ \emph {et~al.}(2012)\citenamefont {Fowler}, \citenamefont {Mariantoni}, \citenamefont {Martinis},\ and\ \citenamefont {Cleland}}]{Fowler2012}%
  \BibitemOpen
  \bibfield  {author} {\bibinfo {author} {\bibfnamefont {A.~G.}\ \bibnamefont {Fowler}}, \bibinfo {author} {\bibfnamefont {M.}~\bibnamefont {Mariantoni}}, \bibinfo {author} {\bibfnamefont {J.~M.}\ \bibnamefont {Martinis}},\ and\ \bibinfo {author} {\bibfnamefont {A.~N.}\ \bibnamefont {Cleland}},\ }\bibfield  {title} {\bibinfo {title} {Surface codes: Towards practical large-scale quantum computation},\ }\href {https://doi.org/10.1103/PhysRevA.86.032324} {\bibfield  {journal} {\bibinfo  {journal} {Phys. Rev. A}\ }\textbf {\bibinfo {volume} {86}},\ \bibinfo {pages} {032324} (\bibinfo {year} {2012})}\BibitemShut {NoStop}%
\bibitem [{\citenamefont {Gonzalez-Zalba}\ \emph {et~al.}(2021)\citenamefont {Gonzalez-Zalba}, \citenamefont {de~Franceschi}, \citenamefont {Charbon}, \citenamefont {Meunier}, \citenamefont {Vinet},\ and\ \citenamefont {Dzurak}}]{fernando2021}%
  \BibitemOpen
  \bibfield  {author} {\bibinfo {author} {\bibfnamefont {M.~F.}\ \bibnamefont {Gonzalez-Zalba}}, \bibinfo {author} {\bibfnamefont {S.}~\bibnamefont {de~Franceschi}}, \bibinfo {author} {\bibfnamefont {E.}~\bibnamefont {Charbon}}, \bibinfo {author} {\bibfnamefont {T.}~\bibnamefont {Meunier}}, \bibinfo {author} {\bibfnamefont {M.}~\bibnamefont {Vinet}},\ and\ \bibinfo {author} {\bibfnamefont {A.~S.}\ \bibnamefont {Dzurak}},\ }\bibfield  {title} {\bibinfo {title} {Scaling silicon-based quantum computing using {CMOS} technology},\ }\href {https://doi.org/10.1038/s41928-021-00681-y} {\bibfield  {journal} {\bibinfo  {journal} {Nature Electronics}\ }\textbf {\bibinfo {volume} {4}},\ \bibinfo {pages} {872} (\bibinfo {year} {2021})}\BibitemShut {NoStop}%
\bibitem [{\citenamefont {Xue}\ \emph {et~al.}(2021)\citenamefont {Xue}, \citenamefont {Patra}, \citenamefont {van Dijk}, \citenamefont {Samkharadze}, \citenamefont {Subramanian}, \citenamefont {Corna}, \citenamefont {Paquelet~Wuetz}, \citenamefont {Jeon}, \citenamefont {Sheikh}, \citenamefont {Juarez-Hernandez}, \citenamefont {Esparza}, \citenamefont {Rampurawala}, \citenamefont {Carlton}, \citenamefont {Ravikumar}, \citenamefont {Nieva}, \citenamefont {Kim}, \citenamefont {Lee}, \citenamefont {Sammak}, \citenamefont {Scappucci}, \citenamefont {Veldhorst}, \citenamefont {Sebastiano}, \citenamefont {Babaie}, \citenamefont {Pellerano}, \citenamefont {Charbon},\ and\ \citenamefont {Vandersypen}}]{Xue2021}%
  \BibitemOpen
  \bibfield  {author} {\bibinfo {author} {\bibfnamefont {X.}~\bibnamefont {Xue}}, \bibinfo {author} {\bibfnamefont {B.}~\bibnamefont {Patra}}, \bibinfo {author} {\bibfnamefont {J.~P.~G.}\ \bibnamefont {van Dijk}}, \bibinfo {author} {\bibfnamefont {N.}~\bibnamefont {Samkharadze}}, \bibinfo {author} {\bibfnamefont {S.}~\bibnamefont {Subramanian}}, \bibinfo {author} {\bibfnamefont {A.}~\bibnamefont {Corna}}, \bibinfo {author} {\bibfnamefont {B.}~\bibnamefont {Paquelet~Wuetz}}, \bibinfo {author} {\bibfnamefont {C.}~\bibnamefont {Jeon}}, \bibinfo {author} {\bibfnamefont {F.}~\bibnamefont {Sheikh}}, \bibinfo {author} {\bibfnamefont {E.}~\bibnamefont {Juarez-Hernandez}}, \bibinfo {author} {\bibfnamefont {B.~P.}\ \bibnamefont {Esparza}}, \bibinfo {author} {\bibfnamefont {H.}~\bibnamefont {Rampurawala}}, \bibinfo {author} {\bibfnamefont {B.}~\bibnamefont {Carlton}}, \bibinfo {author} {\bibfnamefont {S.}~\bibnamefont {Ravikumar}}, \bibinfo {author} {\bibfnamefont {C.}~\bibnamefont {Nieva}}, \bibinfo {author}
  {\bibfnamefont {S.}~\bibnamefont {Kim}}, \bibinfo {author} {\bibfnamefont {H.-J.}\ \bibnamefont {Lee}}, \bibinfo {author} {\bibfnamefont {A.}~\bibnamefont {Sammak}}, \bibinfo {author} {\bibfnamefont {G.}~\bibnamefont {Scappucci}}, \bibinfo {author} {\bibfnamefont {M.}~\bibnamefont {Veldhorst}}, \bibinfo {author} {\bibfnamefont {F.}~\bibnamefont {Sebastiano}}, \bibinfo {author} {\bibfnamefont {M.}~\bibnamefont {Babaie}}, \bibinfo {author} {\bibfnamefont {S.}~\bibnamefont {Pellerano}}, \bibinfo {author} {\bibfnamefont {E.}~\bibnamefont {Charbon}},\ and\ \bibinfo {author} {\bibfnamefont {L.~M.~K.}\ \bibnamefont {Vandersypen}},\ }\bibfield  {title} {\bibinfo {title} {{CMOS}-based cryogenic control of silicon quantum circuits},\ }\href {https://doi.org/10.1038/s41586-021-03469-4} {\bibfield  {journal} {\bibinfo  {journal} {Nature}\ }\textbf {\bibinfo {volume} {593}},\ \bibinfo {pages} {205} (\bibinfo {year} {2021})}\BibitemShut {NoStop}%
\bibitem [{\citenamefont {Paradkar}\ \emph {et~al.}(2025)\citenamefont {Paradkar}, \citenamefont {Nicaise}, \citenamefont {Dakroury}, \citenamefont {Resare},\ and\ \citenamefont {Wieczorek}}]{Paradkar2025FlipChipIndium}%
  \BibitemOpen
  \bibfield  {author} {\bibinfo {author} {\bibfnamefont {A.}~\bibnamefont {Paradkar}}, \bibinfo {author} {\bibfnamefont {P.}~\bibnamefont {Nicaise}}, \bibinfo {author} {\bibfnamefont {K.}~\bibnamefont {Dakroury}}, \bibinfo {author} {\bibfnamefont {F.}~\bibnamefont {Resare}},\ and\ \bibinfo {author} {\bibfnamefont {W.}~\bibnamefont {Wieczorek}},\ }\bibfield  {title} {\bibinfo {title} {Superconducting flip-chip devices using indium microspheres on au-passivated {N}b or {N}b{N} as under-bump metallization layer},\ }\href {https://doi.org/10.1063/5.0235266} {\bibfield  {journal} {\bibinfo  {journal} {Applied Physics Letters}\ }\textbf {\bibinfo {volume} {126}},\ \bibinfo {pages} {022601} (\bibinfo {year} {2025})}\BibitemShut {NoStop}%
\bibitem [{\citenamefont {Bartee}\ \emph {et~al.}(2025)\citenamefont {Bartee}, \citenamefont {Gilbert}, \citenamefont {Zuo}, \citenamefont {Das}, \citenamefont {Tanttu}, \citenamefont {Yang}, \citenamefont {Stuyck}, \citenamefont {Pauka}, \citenamefont {Su}, \citenamefont {Lim}, \citenamefont {Serrano}, \citenamefont {Escott}, \citenamefont {Hudson}, \citenamefont {Itoh}, \citenamefont {Laucht}, \citenamefont {Dzurak},\ and\ \citenamefont {Reilly}}]{Bartee2025}%
  \BibitemOpen
  \bibfield  {author} {\bibinfo {author} {\bibfnamefont {S.~K.}\ \bibnamefont {Bartee}}, \bibinfo {author} {\bibfnamefont {W.}~\bibnamefont {Gilbert}}, \bibinfo {author} {\bibfnamefont {K.}~\bibnamefont {Zuo}}, \bibinfo {author} {\bibfnamefont {K.}~\bibnamefont {Das}}, \bibinfo {author} {\bibfnamefont {T.}~\bibnamefont {Tanttu}}, \bibinfo {author} {\bibfnamefont {C.~H.}\ \bibnamefont {Yang}}, \bibinfo {author} {\bibfnamefont {N.~D.}\ \bibnamefont {Stuyck}}, \bibinfo {author} {\bibfnamefont {S.~J.}\ \bibnamefont {Pauka}}, \bibinfo {author} {\bibfnamefont {R.~Y.}\ \bibnamefont {Su}}, \bibinfo {author} {\bibfnamefont {W.~H.}\ \bibnamefont {Lim}}, \bibinfo {author} {\bibfnamefont {S.}~\bibnamefont {Serrano}}, \bibinfo {author} {\bibfnamefont {C.~C.}\ \bibnamefont {Escott}}, \bibinfo {author} {\bibfnamefont {F.~E.}\ \bibnamefont {Hudson}}, \bibinfo {author} {\bibfnamefont {K.~M.}\ \bibnamefont {Itoh}}, \bibinfo {author} {\bibfnamefont {A.}~\bibnamefont {Laucht}}, \bibinfo {author} {\bibfnamefont {A.~S.}\
  \bibnamefont {Dzurak}},\ and\ \bibinfo {author} {\bibfnamefont {D.~J.}\ \bibnamefont {Reilly}},\ }\bibfield  {title} {\bibinfo {title} {Spin-qubit control with a milli-kelvin {CMOS} chip},\ }\href {https://doi.org/10.1038/s41586-025-09157-x} {\bibfield  {journal} {\bibinfo  {journal} {Nature}\ }\textbf {\bibinfo {volume} {643}},\ \bibinfo {pages} {382} (\bibinfo {year} {2025})}\BibitemShut {NoStop}%
\bibitem [{\citenamefont {King}\ \emph {et~al.}(2023)\citenamefont {King}, \citenamefont {Raymond}, \citenamefont {Lanting}, \citenamefont {Harris}, \citenamefont {Zucca}, \citenamefont {Altomare}, \citenamefont {Berkley}, \citenamefont {Boothby}, \citenamefont {Ejtemaee}, \citenamefont {Enderud}, \citenamefont {Hoskinson}, \citenamefont {Huang}, \citenamefont {Ladizinsky}, \citenamefont {MacDonald}, \citenamefont {Marsden}, \citenamefont {Molavi}, \citenamefont {Oh}, \citenamefont {Poulin-Lamarre}, \citenamefont {Reis}, \citenamefont {Rich}, \citenamefont {Sato}, \citenamefont {Tsai}, \citenamefont {Volkmann}, \citenamefont {Whittaker},\ and\ \citenamefont {Amin}}]{King2023QuantumCritical5000}%
  \BibitemOpen
  \bibfield  {author} {\bibinfo {author} {\bibfnamefont {A.~D.}\ \bibnamefont {King}}, \bibinfo {author} {\bibfnamefont {J.}~\bibnamefont {Raymond}}, \bibinfo {author} {\bibfnamefont {T.}~\bibnamefont {Lanting}}, \bibinfo {author} {\bibfnamefont {R.}~\bibnamefont {Harris}}, \bibinfo {author} {\bibfnamefont {A.}~\bibnamefont {Zucca}}, \bibinfo {author} {\bibfnamefont {F.}~\bibnamefont {Altomare}}, \bibinfo {author} {\bibfnamefont {A.~J.}\ \bibnamefont {Berkley}}, \bibinfo {author} {\bibfnamefont {K.}~\bibnamefont {Boothby}}, \bibinfo {author} {\bibfnamefont {S.}~\bibnamefont {Ejtemaee}}, \bibinfo {author} {\bibfnamefont {C.}~\bibnamefont {Enderud}}, \bibinfo {author} {\bibfnamefont {E.}~\bibnamefont {Hoskinson}}, \bibinfo {author} {\bibfnamefont {S.}~\bibnamefont {Huang}}, \bibinfo {author} {\bibfnamefont {E.}~\bibnamefont {Ladizinsky}}, \bibinfo {author} {\bibfnamefont {A.~J.~R.}\ \bibnamefont {MacDonald}}, \bibinfo {author} {\bibfnamefont {G.}~\bibnamefont {Marsden}}, \bibinfo {author} {\bibfnamefont
  {R.}~\bibnamefont {Molavi}}, \bibinfo {author} {\bibfnamefont {T.}~\bibnamefont {Oh}}, \bibinfo {author} {\bibfnamefont {G.}~\bibnamefont {Poulin-Lamarre}}, \bibinfo {author} {\bibfnamefont {M.}~\bibnamefont {Reis}}, \bibinfo {author} {\bibfnamefont {C.}~\bibnamefont {Rich}}, \bibinfo {author} {\bibfnamefont {Y.}~\bibnamefont {Sato}}, \bibinfo {author} {\bibfnamefont {N.}~\bibnamefont {Tsai}}, \bibinfo {author} {\bibfnamefont {M.}~\bibnamefont {Volkmann}}, \bibinfo {author} {\bibfnamefont {J.~D.}\ \bibnamefont {Whittaker}},\ and\ \bibinfo {author} {\bibfnamefont {M.~H.}\ \bibnamefont {Amin}},\ }\bibfield  {title} {\bibinfo {title} {Quantum critical dynamics in a 5,000-qubit programmable spin glass},\ }\href {https://doi.org/10.1038/s41586-023-05867-2} {\bibfield  {journal} {\bibinfo  {journal} {Nature}\ }\textbf {\bibinfo {volume} {617}},\ \bibinfo {pages} {61} (\bibinfo {year} {2023})}\BibitemShut {NoStop}%
\bibitem [{\citenamefont {Reilly}(2019)}]{reilly2019challengesscalingupcontrolinterface}%
  \BibitemOpen
  \bibfield  {author} {\bibinfo {author} {\bibfnamefont {D.~J.}\ \bibnamefont {Reilly}},\ }\href {https://arxiv.org/abs/1912.05114} {\bibinfo {title} {Challenges in scaling-up the control interface of a quantum computer}} (\bibinfo {year} {2019}),\ \url{https://arxiv.org/abs/1912.05114} {arXiv:1912.05114 [quant-ph]} \BibitemShut {NoStop}%
\bibitem [{\citenamefont {Yoneda}\ \emph {et~al.}(2020)\citenamefont {Yoneda}, \citenamefont {Takeda}, \citenamefont {Noiri}, \citenamefont {Nakajima}, \citenamefont {Li}, \citenamefont {Kamioka}, \citenamefont {Kodera},\ and\ \citenamefont {Tarucha}}]{Yoneda2020}%
  \BibitemOpen
  \bibfield  {author} {\bibinfo {author} {\bibfnamefont {J.}~\bibnamefont {Yoneda}}, \bibinfo {author} {\bibfnamefont {K.}~\bibnamefont {Takeda}}, \bibinfo {author} {\bibfnamefont {A.}~\bibnamefont {Noiri}}, \bibinfo {author} {\bibfnamefont {T.}~\bibnamefont {Nakajima}}, \bibinfo {author} {\bibfnamefont {S.}~\bibnamefont {Li}}, \bibinfo {author} {\bibfnamefont {J.}~\bibnamefont {Kamioka}}, \bibinfo {author} {\bibfnamefont {T.}~\bibnamefont {Kodera}},\ and\ \bibinfo {author} {\bibfnamefont {S.}~\bibnamefont {Tarucha}},\ }\bibfield  {title} {\bibinfo {title} {Quantum non-demolition readout of an electron spin in silicon},\ }\bibfield  {journal} {\bibinfo  {journal} {Nature Communications}\ }\textbf {\bibinfo {volume} {11}},\ \href {https://doi.org/10.1038/s41467-020-14818-8} {10.1038/s41467-020-14818-8} (\bibinfo {year} {2020})\BibitemShut {NoStop}%
\bibitem [{\citenamefont {Blumoff}\ \emph {et~al.}(2022)\citenamefont {Blumoff}, \citenamefont {Pan}, \citenamefont {Keating}, \citenamefont {Andrews}, \citenamefont {Barnes}, \citenamefont {Brecht}, \citenamefont {Croke}, \citenamefont {Euliss}, \citenamefont {Fast}, \citenamefont {Jackson}, \citenamefont {Jones}, \citenamefont {Kerckhoff}, \citenamefont {Lanza}, \citenamefont {Raach}, \citenamefont {Thomas}, \citenamefont {Velunta}, \citenamefont {Weinstein}, \citenamefont {Ladd}, \citenamefont {Eng}, \citenamefont {Borselli}, \citenamefont {Hunter},\ and\ \citenamefont {Rakher}}]{PRXQuantum.3.010352}%
  \BibitemOpen
  \bibfield  {author} {\bibinfo {author} {\bibfnamefont {J.~Z.}\ \bibnamefont {Blumoff}}, \bibinfo {author} {\bibfnamefont {A.~S.}\ \bibnamefont {Pan}}, \bibinfo {author} {\bibfnamefont {T.~E.}\ \bibnamefont {Keating}}, \bibinfo {author} {\bibfnamefont {R.~W.}\ \bibnamefont {Andrews}}, \bibinfo {author} {\bibfnamefont {D.~W.}\ \bibnamefont {Barnes}}, \bibinfo {author} {\bibfnamefont {T.~L.}\ \bibnamefont {Brecht}}, \bibinfo {author} {\bibfnamefont {E.~T.}\ \bibnamefont {Croke}}, \bibinfo {author} {\bibfnamefont {L.~E.}\ \bibnamefont {Euliss}}, \bibinfo {author} {\bibfnamefont {J.~A.}\ \bibnamefont {Fast}}, \bibinfo {author} {\bibfnamefont {C.~A.}\ \bibnamefont {Jackson}}, \bibinfo {author} {\bibfnamefont {A.~M.}\ \bibnamefont {Jones}}, \bibinfo {author} {\bibfnamefont {J.}~\bibnamefont {Kerckhoff}}, \bibinfo {author} {\bibfnamefont {R.~K.}\ \bibnamefont {Lanza}}, \bibinfo {author} {\bibfnamefont {K.}~\bibnamefont {Raach}}, \bibinfo {author} {\bibfnamefont {B.~J.}\ \bibnamefont {Thomas}}, \bibinfo {author}
  {\bibfnamefont {R.}~\bibnamefont {Velunta}}, \bibinfo {author} {\bibfnamefont {A.~J.}\ \bibnamefont {Weinstein}}, \bibinfo {author} {\bibfnamefont {T.~D.}\ \bibnamefont {Ladd}}, \bibinfo {author} {\bibfnamefont {K.}~\bibnamefont {Eng}}, \bibinfo {author} {\bibfnamefont {M.~G.}\ \bibnamefont {Borselli}}, \bibinfo {author} {\bibfnamefont {A.~T.}\ \bibnamefont {Hunter}},\ and\ \bibinfo {author} {\bibfnamefont {M.~T.}\ \bibnamefont {Rakher}},\ }\bibfield  {title} {\bibinfo {title} {Fast and high-fidelity state preparation and measurement in triple-quantum-dot spin qubits},\ }\href {https://doi.org/10.1103/PRXQuantum.3.010352} {\bibfield  {journal} {\bibinfo  {journal} {PRX Quantum}\ }\textbf {\bibinfo {volume} {3}},\ \bibinfo {pages} {010352} (\bibinfo {year} {2022})}\BibitemShut {NoStop}%
\bibitem [{\citenamefont {Mills}\ \emph {et~al.}(2022)\citenamefont {Mills}, \citenamefont {Guinn}, \citenamefont {Gullans}, \citenamefont {Sigillito}, \citenamefont {Feldman}, \citenamefont {Nielsen},\ and\ \citenamefont {Petta}}]{Mills2022}%
  \BibitemOpen
  \bibfield  {author} {\bibinfo {author} {\bibfnamefont {A.~R.}\ \bibnamefont {Mills}}, \bibinfo {author} {\bibfnamefont {C.~R.}\ \bibnamefont {Guinn}}, \bibinfo {author} {\bibfnamefont {M.~J.}\ \bibnamefont {Gullans}}, \bibinfo {author} {\bibfnamefont {A.~J.}\ \bibnamefont {Sigillito}}, \bibinfo {author} {\bibfnamefont {M.~M.}\ \bibnamefont {Feldman}}, \bibinfo {author} {\bibfnamefont {E.}~\bibnamefont {Nielsen}},\ and\ \bibinfo {author} {\bibfnamefont {J.~R.}\ \bibnamefont {Petta}},\ }\bibfield  {title} {\bibinfo {title} {Two-qubit silicon quantum processor with operation fidelity exceeding 99\%},\ }\href {https://doi.org/10.1126/sciadv.abn5130} {\bibfield  {journal} {\bibinfo  {journal} {Science Advances}\ }\textbf {\bibinfo {volume} {8}},\ \bibinfo {pages} {eabn5130} (\bibinfo {year} {2022})},
  \BibitemShut {NoStop}%
\bibitem [{\citenamefont {Yoneda}\ \emph {et~al.}(2018)\citenamefont {Yoneda}, \citenamefont {Takeda}, \citenamefont {Otsuka}, \citenamefont {Nakajima}, \citenamefont {Delbecq}, \citenamefont {Allison}, \citenamefont {Honda}, \citenamefont {Kodera}, \citenamefont {Oda}, \citenamefont {Hoshi}, \citenamefont {Usami}, \citenamefont {Itoh},\ and\ \citenamefont {Tarucha}}]{Yoneda2018}%
  \BibitemOpen
  \bibfield  {author} {\bibinfo {author} {\bibfnamefont {J.}~\bibnamefont {Yoneda}}, \bibinfo {author} {\bibfnamefont {K.}~\bibnamefont {Takeda}}, \bibinfo {author} {\bibfnamefont {T.}~\bibnamefont {Otsuka}}, \bibinfo {author} {\bibfnamefont {T.}~\bibnamefont {Nakajima}}, \bibinfo {author} {\bibfnamefont {M.~R.}\ \bibnamefont {Delbecq}}, \bibinfo {author} {\bibfnamefont {G.}~\bibnamefont {Allison}}, \bibinfo {author} {\bibfnamefont {T.}~\bibnamefont {Honda}}, \bibinfo {author} {\bibfnamefont {T.}~\bibnamefont {Kodera}}, \bibinfo {author} {\bibfnamefont {S.}~\bibnamefont {Oda}}, \bibinfo {author} {\bibfnamefont {Y.}~\bibnamefont {Hoshi}}, \bibinfo {author} {\bibfnamefont {N.}~\bibnamefont {Usami}}, \bibinfo {author} {\bibfnamefont {K.~M.}\ \bibnamefont {Itoh}},\ and\ \bibinfo {author} {\bibfnamefont {S.}~\bibnamefont {Tarucha}},\ }\bibfield  {title} {\bibinfo {title} {A quantum-dot spin qubit with coherence limited by charge noise and fidelity higher than 99.9\%},\ }\href
  {https://doi.org/10.1038/s41565-017-0014-x} {\bibfield  {journal} {\bibinfo  {journal} {Nature Nanotechnology}\ }\textbf {\bibinfo {volume} {13}},\ \bibinfo {pages} {102} (\bibinfo {year} {2018})}\BibitemShut {NoStop}%
\bibitem [{\citenamefont {Noiri}\ \emph {et~al.}(2022)\citenamefont {Noiri}, \citenamefont {Takeda}, \citenamefont {Nakajima}, \citenamefont {Kobayashi}, \citenamefont {Sammak}, \citenamefont {Scappucci},\ and\ \citenamefont {Tarucha}}]{Noiri2022}%
  \BibitemOpen
  \bibfield  {author} {\bibinfo {author} {\bibfnamefont {A.}~\bibnamefont {Noiri}}, \bibinfo {author} {\bibfnamefont {K.}~\bibnamefont {Takeda}}, \bibinfo {author} {\bibfnamefont {T.}~\bibnamefont {Nakajima}}, \bibinfo {author} {\bibfnamefont {T.}~\bibnamefont {Kobayashi}}, \bibinfo {author} {\bibfnamefont {A.}~\bibnamefont {Sammak}}, \bibinfo {author} {\bibfnamefont {G.}~\bibnamefont {Scappucci}},\ and\ \bibinfo {author} {\bibfnamefont {S.}~\bibnamefont {Tarucha}},\ }\bibfield  {title} {\bibinfo {title} {Fast universal quantum gate above the fault-tolerance threshold in silicon},\ }\href {https://doi.org/10.1038/s41586-021-04182-y} {\bibfield  {journal} {\bibinfo  {journal} {Nature}\ }\textbf {\bibinfo {volume} {601}},\ \bibinfo {pages} {338} (\bibinfo {year} {2022})}\BibitemShut {NoStop}%
\bibitem [{\citenamefont {Steinacker}\ \emph {et~al.}(2024)\citenamefont {Steinacker}, \citenamefont {Stuyck}, \citenamefont {Lim}, \citenamefont {Tanttu}, \citenamefont {Feng}, \citenamefont {Nickl}, \citenamefont {Serrano}, \citenamefont {Candido}, \citenamefont {Cifuentes}, \citenamefont {Hudson}, \citenamefont {Chan}, \citenamefont {Kubicek}, \citenamefont {Jussot}, \citenamefont {Canvel}, \citenamefont {Beyne}, \citenamefont {Shimura}, \citenamefont {Loo}, \citenamefont {Godfrin}, \citenamefont {Raes}, \citenamefont {Baudot}, \citenamefont {Wan}, \citenamefont {Laucht}, \citenamefont {Yang}, \citenamefont {Saraiva}, \citenamefont {Escott}, \citenamefont {Greve},\ and\ \citenamefont {Dzurak}}]{steinacker2024}%
  \BibitemOpen
  \bibfield  {author} {\bibinfo {author} {\bibfnamefont {P.}~\bibnamefont {Steinacker}}, \bibinfo {author} {\bibfnamefont {N.~D.}\ \bibnamefont {Stuyck}}, \bibinfo {author} {\bibfnamefont {W.~H.}\ \bibnamefont {Lim}}, \bibinfo {author} {\bibfnamefont {T.}~\bibnamefont {Tanttu}}, \bibinfo {author} {\bibfnamefont {M.}~\bibnamefont {Feng}}, \bibinfo {author} {\bibfnamefont {A.}~\bibnamefont {Nickl}}, \bibinfo {author} {\bibfnamefont {S.}~\bibnamefont {Serrano}}, \bibinfo {author} {\bibfnamefont {M.}~\bibnamefont {Candido}}, \bibinfo {author} {\bibfnamefont {J.~D.}\ \bibnamefont {Cifuentes}}, \bibinfo {author} {\bibfnamefont {F.~E.}\ \bibnamefont {Hudson}}, \bibinfo {author} {\bibfnamefont {K.~W.}\ \bibnamefont {Chan}}, \bibinfo {author} {\bibfnamefont {S.}~\bibnamefont {Kubicek}}, \bibinfo {author} {\bibfnamefont {J.}~\bibnamefont {Jussot}}, \bibinfo {author} {\bibfnamefont {Y.}~\bibnamefont {Canvel}}, \bibinfo {author} {\bibfnamefont {S.}~\bibnamefont {Beyne}}, \bibinfo {author} {\bibfnamefont {Y.}~\bibnamefont
  {Shimura}}, \bibinfo {author} {\bibfnamefont {R.}~\bibnamefont {Loo}}, \bibinfo {author} {\bibfnamefont {C.}~\bibnamefont {Godfrin}}, \bibinfo {author} {\bibfnamefont {B.}~\bibnamefont {Raes}}, \bibinfo {author} {\bibfnamefont {S.}~\bibnamefont {Baudot}}, \bibinfo {author} {\bibfnamefont {D.}~\bibnamefont {Wan}}, \bibinfo {author} {\bibfnamefont {A.}~\bibnamefont {Laucht}}, \bibinfo {author} {\bibfnamefont {C.~H.}\ \bibnamefont {Yang}}, \bibinfo {author} {\bibfnamefont {A.}~\bibnamefont {Saraiva}}, \bibinfo {author} {\bibfnamefont {C.~C.}\ \bibnamefont {Escott}}, \bibinfo {author} {\bibfnamefont {K.~D.}\ \bibnamefont {Greve}},\ and\ \bibinfo {author} {\bibfnamefont {A.~S.}\ \bibnamefont {Dzurak}},\ }\bibfield  {title} {\bibinfo {title} {A 300 mm foundry silicon spin qubit unit cell exceeding 99\% fidelity in all operations},\ }\href {https://doi.org/10.1038/s41586-025-09531-9} {\bibfield  {journal} {\bibinfo  {journal} {Nature}\ }\textbf {\bibinfo {volume} {646}},\ \bibinfo {pages} {81-87} (\bibinfo {year} {2025})}\BibitemShut {NoStop}%%
  \BibitemShut {NoStop}%
\bibitem [{\citenamefont {Zwerver}\ \emph {et~al.}(2022)\citenamefont {Zwerver}, \citenamefont {Krähenmann}, \citenamefont {Watson}, \citenamefont {Lampert}, \citenamefont {George}, \citenamefont {Pillarisetty}, \citenamefont {Bojarski}, \citenamefont {Amin}, \citenamefont {Amitonov}, \citenamefont {Boter}, \citenamefont {Caudillo}, \citenamefont {Corras-Serrano}, \citenamefont {Dehollain}, \citenamefont {Droulers}, \citenamefont {Henry}, \citenamefont {Kotlyar}, \citenamefont {Lodari}, \citenamefont {Lüthi}, \citenamefont {Michalak}, \citenamefont {Mueller}, \citenamefont {Neyens}, \citenamefont {Roberts}, \citenamefont {Samkharadze}, \citenamefont {Zheng}, \citenamefont {Zietz}, \citenamefont {Scappucci}, \citenamefont {Veldhorst}, \citenamefont {Vandersypen},\ and\ \citenamefont {Clarke}}]{Zwerver2022}%
  \BibitemOpen
  \bibfield  {author} {\bibinfo {author} {\bibfnamefont {A.~M.}\ \bibnamefont {Zwerver}}, \bibinfo {author} {\bibfnamefont {T.}~\bibnamefont {Krähenmann}}, \bibinfo {author} {\bibfnamefont {T.~F.}\ \bibnamefont {Watson}}, \bibinfo {author} {\bibfnamefont {L.}~\bibnamefont {Lampert}}, \bibinfo {author} {\bibfnamefont {H.~C.}\ \bibnamefont {George}}, \bibinfo {author} {\bibfnamefont {R.}~\bibnamefont {Pillarisetty}}, \bibinfo {author} {\bibfnamefont {S.~A.}\ \bibnamefont {Bojarski}}, \bibinfo {author} {\bibfnamefont {P.}~\bibnamefont {Amin}}, \bibinfo {author} {\bibfnamefont {S.~V.}\ \bibnamefont {Amitonov}}, \bibinfo {author} {\bibfnamefont {J.~M.}\ \bibnamefont {Boter}}, \bibinfo {author} {\bibfnamefont {R.}~\bibnamefont {Caudillo}}, \bibinfo {author} {\bibfnamefont {D.}~\bibnamefont {Corras-Serrano}}, \bibinfo {author} {\bibfnamefont {J.~P.}\ \bibnamefont {Dehollain}}, \bibinfo {author} {\bibfnamefont {G.}~\bibnamefont {Droulers}}, \bibinfo {author} {\bibfnamefont {E.~M.}\ \bibnamefont {Henry}}, \bibinfo
  {author} {\bibfnamefont {R.}~\bibnamefont {Kotlyar}}, \bibinfo {author} {\bibfnamefont {M.}~\bibnamefont {Lodari}}, \bibinfo {author} {\bibfnamefont {F.}~\bibnamefont {Lüthi}}, \bibinfo {author} {\bibfnamefont {D.~J.}\ \bibnamefont {Michalak}}, \bibinfo {author} {\bibfnamefont {B.~K.}\ \bibnamefont {Mueller}}, \bibinfo {author} {\bibfnamefont {S.}~\bibnamefont {Neyens}}, \bibinfo {author} {\bibfnamefont {J.}~\bibnamefont {Roberts}}, \bibinfo {author} {\bibfnamefont {N.}~\bibnamefont {Samkharadze}}, \bibinfo {author} {\bibfnamefont {G.}~\bibnamefont {Zheng}}, \bibinfo {author} {\bibfnamefont {O.~K.}\ \bibnamefont {Zietz}}, \bibinfo {author} {\bibfnamefont {G.}~\bibnamefont {Scappucci}}, \bibinfo {author} {\bibfnamefont {M.}~\bibnamefont {Veldhorst}}, \bibinfo {author} {\bibfnamefont {L.~M.}\ \bibnamefont {Vandersypen}},\ and\ \bibinfo {author} {\bibfnamefont {J.~S.}\ \bibnamefont {Clarke}},\ }\bibfield  {title} {\bibinfo {title} {Qubits made by advanced semiconductor manufacturing},\ }\href
  {https://doi.org/10.1038/s41928-022-00727-9} {\bibfield  {journal} {\bibinfo  {journal} {Nature Electronics}\ }\textbf {\bibinfo {volume} {5}},\ \bibinfo {pages} {184} (\bibinfo {year} {2022})}\BibitemShut {NoStop}%
\bibitem [{\citenamefont {Chittock-Wood}\ \emph {et~al.}(2025)\citenamefont {Chittock-Wood}, \citenamefont {Leon}, \citenamefont {Fogarty}, \citenamefont {Murphy}, \citenamefont {Patomäki}, \citenamefont {Oakes}, \citenamefont {Williams}, \citenamefont {von Horstig}, \citenamefont {Johnson}, \citenamefont {Jussot}, \citenamefont {Kubicek}, \citenamefont {Govoreanu}, \citenamefont {Wise}, \citenamefont {Gonzalez-Zalba},\ and\ \citenamefont {Morton}}]{chittockwood2025}%
  \BibitemOpen
  \bibfield  {author} {\bibinfo {author} {\bibfnamefont {J.~F.}\ \bibnamefont {Chittock-Wood}}, \bibinfo {author} {\bibfnamefont {R.~C.~C.}\ \bibnamefont {Leon}}, \bibinfo {author} {\bibfnamefont {M.~A.}\ \bibnamefont {Fogarty}}, \bibinfo {author} {\bibfnamefont {T.}~\bibnamefont {Murphy}}, \bibinfo {author} {\bibfnamefont {S.~M.}\ \bibnamefont {Patomäki}}, \bibinfo {author} {\bibfnamefont {G.~A.}\ \bibnamefont {Oakes}}, \bibinfo {author} {\bibfnamefont {J.}~\bibnamefont {Williams}}, \bibinfo {author} {\bibfnamefont {F.-E.}\ \bibnamefont {von Horstig}}, \bibinfo {author} {\bibfnamefont {N.}~\bibnamefont {Johnson}}, \bibinfo {author} {\bibfnamefont {J.}~\bibnamefont {Jussot}}, \bibinfo {author} {\bibfnamefont {S.}~\bibnamefont {Kubicek}}, \bibinfo {author} {\bibfnamefont {B.}~\bibnamefont {Govoreanu}}, \bibinfo {author} {\bibfnamefont {D.~F.}\ \bibnamefont {Wise}}, \bibinfo {author} {\bibfnamefont {M.~F.}\ \bibnamefont {Gonzalez-Zalba}},\ and\ \bibinfo {author} {\bibfnamefont {J.~J.~L.}\ \bibnamefont
  {Morton}},\ }\href {https://arxiv.org/abs/2408.01241} {\bibinfo {title} {Radio-frequency cascade readout of coupled spin qubits fabricated using a 300~mm wafer process}} (\bibinfo {year} {2025}),
  \BibitemShut {NoStop}%
\bibitem [{\citenamefont {Lainé}\ \emph {et~al.}(2025)\citenamefont {Lainé}, \citenamefont {Oakes}, \citenamefont {Ciriano-Tejel}, \citenamefont {Chittock-Wood}, \citenamefont {Peri}, \citenamefont {Fogarty}, \citenamefont {Patomäki}, \citenamefont {Kubicek}, \citenamefont {Wise}, \citenamefont {Leon}, \citenamefont {Gonzalez-Zalba},\ and\ \citenamefont {Morton}}]{Laine2025}%
  \BibitemOpen
  \bibfield  {author} {\bibinfo {author} {\bibfnamefont {C.}~\bibnamefont {Lainé}}, \bibinfo {author} {\bibfnamefont {G.~A.}\ \bibnamefont {Oakes}}, \bibinfo {author} {\bibfnamefont {V.}~\bibnamefont {Ciriano-Tejel}}, \bibinfo {author} {\bibfnamefont {J.~F.}\ \bibnamefont {Chittock-Wood}}, \bibinfo {author} {\bibfnamefont {L.}~\bibnamefont {Peri}}, \bibinfo {author} {\bibfnamefont {M.~A.}\ \bibnamefont {Fogarty}}, \bibinfo {author} {\bibfnamefont {S.~M.}\ \bibnamefont {Patomäki}}, \bibinfo {author} {\bibfnamefont {S.}~\bibnamefont {Kubicek}}, \bibinfo {author} {\bibfnamefont {D.~F.}\ \bibnamefont {Wise}}, \bibinfo {author} {\bibfnamefont {R.~C.~C.}\ \bibnamefont {Leon}}, \bibinfo {author} {\bibfnamefont {M.~F.}\ \bibnamefont {Gonzalez-Zalba}},\ and\ \bibinfo {author} {\bibfnamefont {J.~J.~L.}\ \bibnamefont {Morton}},\ }\href {https://arxiv.org/abs/2505.10435} {\bibinfo {title} {High-fidelity dispersive spin sensing in a tuneable unit cell of silicon {MOS} quantum dots}} (\bibinfo {year} {2025}), \BibitemShut {NoStop}%
\bibitem [{\citenamefont {Hamonic}\ \emph {et~al.}(2025)\citenamefont {Hamonic}, \citenamefont {Toubeix}, \citenamefont {Haas}, \citenamefont {Nath}, \citenamefont {Dartiailh}, \citenamefont {Martinez}, \citenamefont {Bertrand}, \citenamefont {Niebojewski}, \citenamefont {Vinet}, \citenamefont {Bäuerle}, \citenamefont {Balestro}, \citenamefont {Meunier},\ and\ \citenamefont {Urdampilleta}}]{hamonic2025}%
  \BibitemOpen
  \bibfield  {author} {\bibinfo {author} {\bibfnamefont {P.}~\bibnamefont {Hamonic}}, \bibinfo {author} {\bibfnamefont {M.}~\bibnamefont {Toubeix}}, \bibinfo {author} {\bibfnamefont {G.}~\bibnamefont {Haas}}, \bibinfo {author} {\bibfnamefont {J.}~\bibnamefont {Nath}}, \bibinfo {author} {\bibfnamefont {M.~C.}\ \bibnamefont {Dartiailh}}, \bibinfo {author} {\bibfnamefont {B.}~\bibnamefont {Martinez}}, \bibinfo {author} {\bibfnamefont {B.}~\bibnamefont {Bertrand}}, \bibinfo {author} {\bibfnamefont {H.}~\bibnamefont {Niebojewski}}, \bibinfo {author} {\bibfnamefont {M.}~\bibnamefont {Vinet}}, \bibinfo {author} {\bibfnamefont {C.}~\bibnamefont {Bäuerle}}, \bibinfo {author} {\bibfnamefont {F.}~\bibnamefont {Balestro}}, \bibinfo {author} {\bibfnamefont {T.}~\bibnamefont {Meunier}},\ and\ \bibinfo {author} {\bibfnamefont {M.}~\bibnamefont {Urdampilleta}},\ }\href {https://arxiv.org/abs/2504.20572} {\bibinfo {title} {A foundry-fabricated spin qubit unit cell with in-situ dispersive readout}} (\bibinfo {year} {2025}),\
   \BibitemShut {NoStop}%
\bibitem [{\citenamefont {Ibberson}\ \emph {et~al.}(2024)\citenamefont {Ibberson}, \citenamefont {Kirkman}, \citenamefont {Morton}, \citenamefont {Gonzalez-Zalba},\ and\ \citenamefont {Gomez-Saiz}}]{Ibberson2024}%
  \BibitemOpen
  \bibfield  {author} {\bibinfo {author} {\bibfnamefont {D.~J.}\ \bibnamefont {Ibberson}}, \bibinfo {author} {\bibfnamefont {J.}~\bibnamefont {Kirkman}}, \bibinfo {author} {\bibfnamefont {J.~J.~L.}~\bibnamefont {Morton}}, \bibinfo {author} {\bibfnamefont {M.~F.}~\bibnamefont {Gonzalez-Zalba}}, \ and\ \bibinfo {author} {\bibfnamefont {A.}~\bibnamefont {Gomez-Saiz}},\ }\bibfield  {title} {\bibinfo {title} {A Multi-Module Silicon-On-Insulator Chip Assembly Containing Quantum Dots and Cryogenic Radio-Frequency Readout Electronics},\ }in\ \href {10.1109/ICECS61496.2024.10848854} {\emph {\bibinfo {booktitle}
  {2024 31st IEEE International Conference on Electronics, Circuits and Systems (ICECS)}}}\ (\bibinfo {year} {2024})\ pp.\ \bibinfo {pages} {1--4}\BibitemShut {NoStop}%
\bibitem [{\citenamefont {Pauka}\ \emph {et~al.}(2021)\citenamefont {Pauka}, \citenamefont {Das}, \citenamefont {Kalra}, \citenamefont {Moini}, \citenamefont {Yang}, \citenamefont {Trainer}, \citenamefont {Bousquet}, \citenamefont {Cantaloube}, \citenamefont {Dick}, \citenamefont {Gardner}, \citenamefont {Manfra},\ and\ \citenamefont {Reilly}}]{Pauka2021}%
  \BibitemOpen
  \bibfield  {author} {\bibinfo {author} {\bibfnamefont {S.~J.}\ \bibnamefont {Pauka}}, \bibinfo {author} {\bibfnamefont {K.}~\bibnamefont {Das}}, \bibinfo {author} {\bibfnamefont {R.}~\bibnamefont {Kalra}}, \bibinfo {author} {\bibfnamefont {A.}~\bibnamefont {Moini}}, \bibinfo {author} {\bibfnamefont {Y.}~\bibnamefont {Yang}}, \bibinfo {author} {\bibfnamefont {M.}~\bibnamefont {Trainer}}, \bibinfo {author} {\bibfnamefont {A.}~\bibnamefont {Bousquet}}, \bibinfo {author} {\bibfnamefont {C.}~\bibnamefont {Cantaloube}}, \bibinfo {author} {\bibfnamefont {N.}~\bibnamefont {Dick}}, \bibinfo {author} {\bibfnamefont {G.~C.}\ \bibnamefont {Gardner}}, \bibinfo {author} {\bibfnamefont {M.~J.}\ \bibnamefont {Manfra}},\ and\ \bibinfo {author} {\bibfnamefont {D.~J.}\ \bibnamefont {Reilly}},\ }\bibfield  {title} {\bibinfo {title} {A cryogenic {CMOS} chip for generating control signals for multiple qubits},\ }\href {https://doi.org/10.1038/s41928-020-00528-y} {\bibfield  {journal} {\bibinfo  {journal} {Nature Electronics}\
  }\textbf {\bibinfo {volume} {4}},\ \bibinfo {pages} {64} (\bibinfo {year} {2021})}\BibitemShut {NoStop}%
\bibitem [{\citenamefont {Huang}\ \emph {et~al.}(2024)\citenamefont {Huang}, \citenamefont {Su}, \citenamefont {Lim}, \citenamefont {Feng}, \citenamefont {van Straaten}, \citenamefont {Severin}, \citenamefont {Gilbert}, \citenamefont {Stuyck}, \citenamefont {Tanttu}, \citenamefont {Serrano}, \citenamefont {Cifuentes}, \citenamefont {Hansen}, \citenamefont {Seedhouse}, \citenamefont {Vahapoglu}, \citenamefont {Leon}, \citenamefont {Abrosimov}, \citenamefont {Pohl}, \citenamefont {Thewalt}, \citenamefont {Hudson}, \citenamefont {Escott}, \citenamefont {Ares}, \citenamefont {Bartlett}, \citenamefont {Morello}, \citenamefont {Saraiva}, \citenamefont {Laucht}, \citenamefont {Dzurak},\ and\ \citenamefont {Yang}}]{Huang2024}%
  \BibitemOpen
  \bibfield  {author} {\bibinfo {author} {\bibfnamefont {J.~Y.}\ \bibnamefont {Huang}}, \bibinfo {author} {\bibfnamefont {R.~Y.}\ \bibnamefont {Su}}, \bibinfo {author} {\bibfnamefont {W.~H.}\ \bibnamefont {Lim}}, \bibinfo {author} {\bibfnamefont {M.~K.}\ \bibnamefont {Feng}}, \bibinfo {author} {\bibfnamefont {B.}~\bibnamefont {van Straaten}}, \bibinfo {author} {\bibfnamefont {B.}~\bibnamefont {Severin}}, \bibinfo {author} {\bibfnamefont {W.}~\bibnamefont {Gilbert}}, \bibinfo {author} {\bibfnamefont {N.~D.}\ \bibnamefont {Stuyck}}, \bibinfo {author} {\bibfnamefont {T.}~\bibnamefont {Tanttu}}, \bibinfo {author} {\bibfnamefont {S.}~\bibnamefont {Serrano}}, \bibinfo {author} {\bibfnamefont {J.~D.}\ \bibnamefont {Cifuentes}}, \bibinfo {author} {\bibfnamefont {I.}~\bibnamefont {Hansen}}, \bibinfo {author} {\bibfnamefont {A.~E.}\ \bibnamefont {Seedhouse}}, \bibinfo {author} {\bibfnamefont {E.}~\bibnamefont {Vahapoglu}}, \bibinfo {author} {\bibfnamefont {R.~C.}\ \bibnamefont {Leon}}, \bibinfo {author} {\bibfnamefont
  {N.~V.}\ \bibnamefont {Abrosimov}}, \bibinfo {author} {\bibfnamefont {H.~J.}\ \bibnamefont {Pohl}}, \bibinfo {author} {\bibfnamefont {M.~L.}\ \bibnamefont {Thewalt}}, \bibinfo {author} {\bibfnamefont {F.~E.}\ \bibnamefont {Hudson}}, \bibinfo {author} {\bibfnamefont {C.~C.}\ \bibnamefont {Escott}}, \bibinfo {author} {\bibfnamefont {N.}~\bibnamefont {Ares}}, \bibinfo {author} {\bibfnamefont {S.~D.}\ \bibnamefont {Bartlett}}, \bibinfo {author} {\bibfnamefont {A.}~\bibnamefont {Morello}}, \bibinfo {author} {\bibfnamefont {A.}~\bibnamefont {Saraiva}}, \bibinfo {author} {\bibfnamefont {A.}~\bibnamefont {Laucht}}, \bibinfo {author} {\bibfnamefont {A.~S.}\ \bibnamefont {Dzurak}},\ and\ \bibinfo {author} {\bibfnamefont {C.~H.}\ \bibnamefont {Yang}},\ }\bibfield  {title} {\bibinfo {title} {High-fidelity spin qubit operation and algorithmic initialization above 1\uppercase{K}},\ }\href {https://doi.org/10.1038/s41586-024-07160-2} {\bibfield  {journal} {\bibinfo  {journal} {Nature}\ }\textbf {\bibinfo {volume} {627}},\
  \bibinfo {pages} {772} (\bibinfo {year} {2024})}\BibitemShut {NoStop}%
\bibitem [{\citenamefont {Guevel}\ \emph {et~al.}(2020)\citenamefont {Guevel}, \citenamefont {Billiot}, \citenamefont {Jehl}, \citenamefont {De~Franceschi}, \citenamefont {Zurita}, \citenamefont {Thonnart}, \citenamefont {Vinet}, \citenamefont {Sanquer}, \citenamefont {Maurand}, \citenamefont {Jansen},\ and\ \citenamefont {Pillonnet}}]{Guevel2020}%
  \BibitemOpen
  \bibfield  {author} {\bibinfo {author} {\bibfnamefont {L.~L.}\ \bibnamefont {Guevel}}, \bibinfo {author} {\bibfnamefont {G.}~\bibnamefont {Billiot}}, \bibinfo {author} {\bibfnamefont {X.}~\bibnamefont {Jehl}}, \bibinfo {author} {\bibfnamefont {S.}~\bibnamefont {De~Franceschi}}, \bibinfo {author} {\bibfnamefont {M.}~\bibnamefont {Zurita}}, \bibinfo {author} {\bibfnamefont {Y.}~\bibnamefont {Thonnart}}, \bibinfo {author} {\bibfnamefont {M.}~\bibnamefont {Vinet}}, \bibinfo {author} {\bibfnamefont {M.}~\bibnamefont {Sanquer}}, \bibinfo {author} {\bibfnamefont {R.}~\bibnamefont {Maurand}}, \bibinfo {author} {\bibfnamefont {A.~G.~M.}\ \bibnamefont {Jansen}},\ and\ \bibinfo {author} {\bibfnamefont {G.}~\bibnamefont {Pillonnet}},\ }\bibfield  {title} {\bibinfo {title} {19.2 a 110m{K} 295µ{W} 28nm {FDSOI} {CMOS} quantum integrated circuit with a 2.8{GH}z excitation and n{A} current sensing of an on-chip double quantum dot},\ }in\ \href {https://doi.org/10.1109/ISSCC19947.2020.9063090} {\emph {\bibinfo {booktitle}
  {2020 IEEE International Solid-State Circuits Conference - (ISSCC)}}}\ (\bibinfo {year} {2020})\ pp.\ \bibinfo {pages} {306--308}\BibitemShut {NoStop}%
\bibitem [{\citenamefont {Ruffino}\ \emph {et~al.}(2022)\citenamefont {Ruffino}, \citenamefont {Yang}, \citenamefont {Michniewicz}, \citenamefont {Peng}, \citenamefont {Charbon},\ and\ \citenamefont {Gonzalez-Zalba}}]{Ruffino2022}%
  \BibitemOpen
  \bibfield  {author} {\bibinfo {author} {\bibfnamefont {A.}~\bibnamefont {Ruffino}}, \bibinfo {author} {\bibfnamefont {T.~Y.}\ \bibnamefont {Yang}}, \bibinfo {author} {\bibfnamefont {J.}~\bibnamefont {Michniewicz}}, \bibinfo {author} {\bibfnamefont {Y.}~\bibnamefont {Peng}}, \bibinfo {author} {\bibfnamefont {E.}~\bibnamefont {Charbon}},\ and\ \bibinfo {author} {\bibfnamefont {M.~F.}\ \bibnamefont {Gonzalez-Zalba}},\ }\bibfield  {title} {\bibinfo {title} {A cryo-{CMOS} chip that integrates silicon quantum dots and multiplexed dispersive readout electronics},\ }\href {https://doi.org/10.1038/s41928-021-00687-6} {\bibfield  {journal} {\bibinfo  {journal} {Nature Electronics}\ }\textbf {\bibinfo {volume} {5}},\ \bibinfo {pages} {53} (\bibinfo {year} {2022})}\BibitemShut {NoStop}%
\bibitem [{\citenamefont {Power}\ \emph {et~al.}(2025)\citenamefont {Power}, \citenamefont {Moras}, \citenamefont {Sokolov}, \citenamefont {Rohrbacher}, \citenamefont {Wu}, \citenamefont {Amitonov}, \citenamefont {Kriekouki}, \citenamefont {Aprà}, \citenamefont {Giounanlis}, \citenamefont {Asker}, \citenamefont {Harkin}, \citenamefont {Hanos-Puskai}, \citenamefont {Bisiaux}, \citenamefont {Bashir}, \citenamefont {Redmond}, \citenamefont {Leipold}, \citenamefont {Staszewski}, \citenamefont {Barry}, \citenamefont {Samkharadze},\ and\ \citenamefont {Blokhina}}]{amitonov2024}%
  \BibitemOpen
  \bibfield  {author} {\bibinfo {author} {\bibfnamefont {C.}~\bibnamefont {Power}}, \bibinfo {author} {\bibfnamefont {M.}~\bibnamefont {Moras}}, \bibinfo {author} {\bibfnamefont {A.}~\bibnamefont {Sokolov}}, \bibinfo {author} {\bibfnamefont {C.}~\bibnamefont {Rohrbacher}}, \bibinfo {author} {\bibfnamefont {X.}~\bibnamefont {Wu}}, \bibinfo {author} {\bibfnamefont {S.~V.}\ \bibnamefont {Amitonov}}, \bibinfo {author} {\bibfnamefont {I.}~\bibnamefont {Kriekouki}}, \bibinfo {author} {\bibfnamefont {A.}~\bibnamefont {Aprà}}, \bibinfo {author} {\bibfnamefont {P.}~\bibnamefont {Giounanlis}}, \bibinfo {author} {\bibfnamefont {M.}~\bibnamefont {Asker}}, \bibinfo {author} {\bibfnamefont {M.}~\bibnamefont {Harkin}}, \bibinfo {author} {\bibfnamefont {P.}~\bibnamefont {Hanos-Puskai}}, \bibinfo {author} {\bibfnamefont {P.}~\bibnamefont {Bisiaux}}, \bibinfo {author} {\bibfnamefont {I.}~\bibnamefont {Bashir}}, \bibinfo {author} {\bibfnamefont {D.}~\bibnamefont {Redmond}}, \bibinfo {author} {\bibfnamefont {D.}~\bibnamefont
  {Leipold}}, \bibinfo {author} {\bibfnamefont {R.~B.}\ \bibnamefont {Staszewski}}, \bibinfo {author} {\bibfnamefont {B.}~\bibnamefont {Barry}}, \bibinfo {author} {\bibfnamefont {N.}~\bibnamefont {Samkharadze}},\ and\ \bibinfo {author} {\bibfnamefont {E.}~\bibnamefont {Blokhina}},\ }\bibfield  {title} {\bibinfo {title} {Fully-tunable tunnel-coupled quantum dots and charge sensing in a commercial 22 nm {FD-SOI} process},\ }\href {https://doi.org/10.1109/LED.2025.3595384} {\bibfield  {journal} {\bibinfo  {journal} {IEEE Electron Device Letters}\ }\textbf {\bibinfo {volume} {46}},\ \bibinfo {pages} {1913} (\bibinfo {year} {2025})}\BibitemShut {NoStop}%
\bibitem [{\citenamefont {Thomas}\ \emph {et~al.}(2025)\citenamefont {Thomas}, \citenamefont {Ciriano-Tejel}, \citenamefont {Wise}, \citenamefont {Prete}, \citenamefont {de~Kruijf}, \citenamefont {Ibberson}, \citenamefont {Noah}, \citenamefont {Gomez-Saiz}, \citenamefont {Gonzalez-Zalba}, \citenamefont {Johnson},\ and\ \citenamefont {Morton}}]{Thomas2025}%
  \BibitemOpen
  \bibfield  {author} {\bibinfo {author} {\bibfnamefont {E.~J.}\ \bibnamefont {Thomas}}, \bibinfo {author} {\bibfnamefont {V.~N.}\ \bibnamefont {Ciriano-Tejel}}, \bibinfo {author} {\bibfnamefont {D.~F.}\ \bibnamefont {Wise}}, \bibinfo {author} {\bibfnamefont {D.}~\bibnamefont {Prete}}, \bibinfo {author} {\bibfnamefont {M.}~\bibnamefont {de~Kruijf}}, \bibinfo {author} {\bibfnamefont {D.~J.}\ \bibnamefont {Ibberson}}, \bibinfo {author} {\bibfnamefont {G.~M.}\ \bibnamefont {Noah}}, \bibinfo {author} {\bibfnamefont {A.}~\bibnamefont {Gomez-Saiz}}, \bibinfo {author} {\bibfnamefont {M.~F.}\ \bibnamefont {Gonzalez-Zalba}}, \bibinfo {author} {\bibfnamefont {M.~A.}\ \bibnamefont {Johnson}},\ and\ \bibinfo {author} {\bibfnamefont {J.~J.}\ \bibnamefont {Morton}},\ }\bibfield  {title} {\bibinfo {title} {Rapid cryogenic characterization of 1,024 integrated silicon quantum dot devices},\ }\href {https://doi.org/10.1038/s41928-024-01304-y} {\bibfield  {journal} {\bibinfo  {journal} {Nature Electronics}\ }\textbf {\bibinfo
  {volume} {8}},\ \bibinfo {pages} {75} (\bibinfo {year} {2025})}\BibitemShut {NoStop}%
\bibitem [{\citenamefont {Elzerman}\ \emph {et~al.}(2004)\citenamefont {Elzerman}, \citenamefont {Hanson}, \citenamefont {Willems Van~Beveren}, \citenamefont {Witkamp}, \citenamefont {Vandersypen},\ and\ \citenamefont {Kouwenhoven}}]{elzerman_2004}%
  \BibitemOpen
  \bibfield  {author} {\bibinfo {author} {\bibfnamefont {J.~M.}\ \bibnamefont {Elzerman}}, \bibinfo {author} {\bibfnamefont {R.}~\bibnamefont {Hanson}}, \bibinfo {author} {\bibfnamefont {L.~H.}\ \bibnamefont {Willems Van~Beveren}}, \bibinfo {author} {\bibfnamefont {B.}~\bibnamefont {Witkamp}}, \bibinfo {author} {\bibfnamefont {L.~M.~K.}\ \bibnamefont {Vandersypen}},\ and\ \bibinfo {author} {\bibfnamefont {L.~P.}\ \bibnamefont {Kouwenhoven}},\ }\bibfield  {title} {{\selectlanguage {en}\bibinfo {title} {Single-shot read-out of an individual electron spin in a quantum dot}},\ }\href {https://doi.org/10.1038/nature02693} {\bibfield  {journal} {\bibinfo  {journal} {Nature}\ }\textbf {\bibinfo {volume} {430}},\ \bibinfo {pages} {431} (\bibinfo {year} {2004})}\BibitemShut {NoStop}%
\bibitem [{\citenamefont {Keith}\ \emph {et~al.}(2022)\citenamefont {Keith}, \citenamefont {Chung}, \citenamefont {Kranz}, \citenamefont {Thorgrimsson}, \citenamefont {Gorman},\ and\ \citenamefont {Simmons}}]{Keith2022}%
  \BibitemOpen
  \bibfield  {author} {\bibinfo {author} {\bibfnamefont {D.}~\bibnamefont {Keith}}, \bibinfo {author} {\bibfnamefont {Y.}~\bibnamefont {Chung}}, \bibinfo {author} {\bibfnamefont {L.}~\bibnamefont {Kranz}}, \bibinfo {author} {\bibfnamefont {B.}~\bibnamefont {Thorgrimsson}}, \bibinfo {author} {\bibfnamefont {S.~K.}\ \bibnamefont {Gorman}},\ and\ \bibinfo {author} {\bibfnamefont {M.~Y.}\ \bibnamefont {Simmons}},\ }\bibfield  {title} {\bibinfo {title} {Ramped measurement technique for robust high-fidelity spin qubit readout},\ }\href {https://doi.org/10.1126/sciadv.abq0455} {\bibfield  {journal} {\bibinfo  {journal} {Science Advances}\ }\textbf {\bibinfo {volume} {8}},\ \bibinfo {pages} {eabq0455} (\bibinfo {year} {2022})}\BibitemShut {NoStop}%
\bibitem [{\citenamefont {Wang}\ \emph {et~al.}(2022)\citenamefont {Wang}, \citenamefont {Bi}, \citenamefont {Bu}, \citenamefont {Liu}, \citenamefont {Zhao}, \citenamefont {Cao},\ and\ \citenamefont {Ai}}]{Wang2022}%
  \BibitemOpen
  \bibfield  {author} {\bibinfo {author} {\bibfnamefont {H.}~\bibnamefont {Wang}}, \bibinfo {author} {\bibfnamefont {J.}~\bibnamefont {Bi}}, \bibinfo {author} {\bibfnamefont {J.}~\bibnamefont {Bu}}, \bibinfo {author} {\bibfnamefont {H.}~\bibnamefont {Liu}}, \bibinfo {author} {\bibfnamefont {F.}~\bibnamefont {Zhao}}, \bibinfo {author} {\bibfnamefont {H.}\ \bibnamefont {Cao}},\ and\ \bibinfo {author} {\bibfnamefont {C.}~\bibnamefont {Ai}},\ }\bibfield  {title} {\bibinfo {title} {Characteristics of 22 nm UTBB-FDSOI technology with an ultra-wide temperature range},\ }\href {https://doi.org/10.1088/1361-6641/ac86ec} {\bibfield  {journal} {\bibinfo  {journal} {Semiconductor Science and Technology}\ }\textbf {\bibinfo {volume} {37}},\ \bibinfo {pages} {105004} (\bibinfo {year} {2022})}\BibitemShut {NoStop}%
\bibitem [{\citenamefont {Vigneau}\ \emph {et~al.}(2023)\citenamefont {Vigneau}, \citenamefont {Fedele}, \citenamefont {Chatterjee}, \citenamefont {Reilly}, \citenamefont {Kuemmeth}, \citenamefont {Gonzalez-Zalba}, \citenamefont {Laird},\ and\ \citenamefont {Ares}}]{Vigneau2023}%
  \BibitemOpen
  \bibfield  {author} {\bibinfo {author} {\bibfnamefont {F.}~\bibnamefont {Vigneau}}, \bibinfo {author} {\bibfnamefont {F.}~\bibnamefont {Fedele}}, \bibinfo {author} {\bibfnamefont {A.}~\bibnamefont {Chatterjee}}, \bibinfo {author} {\bibfnamefont {D.}~\bibnamefont {Reilly}}, \bibinfo {author} {\bibfnamefont {F.}~\bibnamefont {Kuemmeth}}, \bibinfo {author} {\bibfnamefont {M.~F.}\ \bibnamefont {Gonzalez-Zalba}}, \bibinfo {author} {\bibfnamefont {E.}~\bibnamefont {Laird}},\ and\ \bibinfo {author} {\bibfnamefont {N.}~\bibnamefont {Ares}},\ }\bibfield  {title} {\bibinfo {title} {Probing quantum devices with radio-frequency reflectometry},\ }\href {https://doi.org/10.1063/5.0088229} {\bibfield  {journal} {\bibinfo  {journal} {Applied Physics Reviews}\ }\textbf {\bibinfo {volume} {10}},\ \bibinfo {pages} {021305} (\bibinfo {year} {2023})}\BibitemShut {NoStop}%
\bibitem [{\citenamefont {Swift}\ \emph {et~al.}(2025)\citenamefont {Swift}, \citenamefont {Olivieri}, \citenamefont {Aizpurua-Iraola}, \citenamefont {Kirkman}, \citenamefont {Noah}, \citenamefont {de~Kruijf}, \citenamefont {von Horstig}, \citenamefont {Gomez-Saiz}, \citenamefont {Morton},\ and\ \citenamefont {Gonzalez-Zalba}}]{swift2025}%
  \BibitemOpen
  \bibfield  {author} {\bibinfo {author} {\bibfnamefont {T.~H.}\ \bibnamefont {Swift}}, \bibinfo {author} {\bibfnamefont {F.}~\bibnamefont {Olivieri}}, \bibinfo {author} {\bibfnamefont {G.}~\bibnamefont {Aizpurua-Iraola}}, \bibinfo {author} {\bibfnamefont {J.}~\bibnamefont {Kirkman}}, \bibinfo {author} {\bibfnamefont {G.~M.}\ \bibnamefont {Noah}}, \bibinfo {author} {\bibfnamefont {M.}~\bibnamefont {de~Kruijf}}, \bibinfo {author} {\bibfnamefont {F.~E.}\ \bibnamefont {von Horstig}}, \bibinfo {author} {\bibfnamefont {A.}~\bibnamefont {Gomez-Saiz}}, \bibinfo {author} {\bibfnamefont {J.~J.~L.}\ \bibnamefont {Morton}},\ and\ \bibinfo {author} {\bibfnamefont {M.~F.}\ \bibnamefont {Gonzalez-Zalba}},\ }\href {https://arxiv.org/abs/2507.13202} {\bibinfo {title} {A superinductor in a deep sub-micron integrated circuit}} (\bibinfo {year} {2025}), \BibitemShut {NoStop}%
\bibitem [{\citenamefont {Morello}\ \emph {et~al.}(2010)\citenamefont {Morello}, \citenamefont {Pla}, \citenamefont {Zwanenburg}, \citenamefont {Chan}, \citenamefont {Tan}, \citenamefont {Huebl}, \citenamefont {M{\"o}tt{\"o}nen}, \citenamefont {Nugroho}, \citenamefont {Yang}, \citenamefont {van Donkelaar}, \citenamefont {Alves}, \citenamefont {Jamieson}, \citenamefont {Escott}, \citenamefont {Hollenberg}, \citenamefont {Clark},\ and\ \citenamefont {Dzurak}}]{Morello2010}%
  \BibitemOpen
  \bibfield  {author} {\bibinfo {author} {\bibfnamefont {A.}~\bibnamefont {Morello}}, \bibinfo {author} {\bibfnamefont {J.~J.}\ \bibnamefont {Pla}}, \bibinfo {author} {\bibfnamefont {F.~A.}\ \bibnamefont {Zwanenburg}}, \bibinfo {author} {\bibfnamefont {K.~W.}\ \bibnamefont {Chan}}, \bibinfo {author} {\bibfnamefont {K.~Y.}\ \bibnamefont {Tan}}, \bibinfo {author} {\bibfnamefont {H.}~\bibnamefont {Huebl}}, \bibinfo {author} {\bibfnamefont {M.}~\bibnamefont {M{\"o}tt{\"o}nen}}, \bibinfo {author} {\bibfnamefont {C.~D.}\ \bibnamefont {Nugroho}}, \bibinfo {author} {\bibfnamefont {C.}~\bibnamefont {Yang}}, \bibinfo {author} {\bibfnamefont {J.~A.}\ \bibnamefont {van Donkelaar}}, \bibinfo {author} {\bibfnamefont {A.~D.~C.}\ \bibnamefont {Alves}}, \bibinfo {author} {\bibfnamefont {D.~N.}\ \bibnamefont {Jamieson}}, \bibinfo {author} {\bibfnamefont {C.~C.}\ \bibnamefont {Escott}}, \bibinfo {author} {\bibfnamefont {L.~C.~L.}\ \bibnamefont {Hollenberg}}, \bibinfo {author} {\bibfnamefont {R.~G.}\ \bibnamefont {Clark}},\ and\
  \bibinfo {author} {\bibfnamefont {A.~S.}\ \bibnamefont {Dzurak}},\ }\bibfield  {title} {\bibinfo {title} {Single-shot readout of an electron spin in silicon},\ }\href {https://doi.org/10.1038/nature09392} {\bibfield  {journal} {\bibinfo  {journal} {Nature}\ }\textbf {\bibinfo {volume} {467}},\ \bibinfo {pages} {687} (\bibinfo {year} {2010})}\BibitemShut {NoStop}%
\bibitem [{\citenamefont {Oakes}\ \emph {et~al.}(2023)\citenamefont {Oakes}, \citenamefont {Ciriano-Tejel}, \citenamefont {Wise}, \citenamefont {Fogarty}, \citenamefont {Lundberg}, \citenamefont {Lain\'e}, \citenamefont {Schaal}, \citenamefont {Martins}, \citenamefont {Ibberson}, \citenamefont {Hutin}, \citenamefont {Bertrand}, \citenamefont {Stelmashenko}, \citenamefont {Robinson}, \citenamefont {Ibberson}, \citenamefont {Hashim}, \citenamefont {Siddiqi}, \citenamefont {Lee}, \citenamefont {Vinet}, \citenamefont {Smith}, \citenamefont {Morton},\ and\ \citenamefont {Gonzalez-Zalba}}]{Oakes2023}%
  \BibitemOpen
  \bibfield  {author} {\bibinfo {author} {\bibfnamefont {G.~A.}\ \bibnamefont {Oakes}}, \bibinfo {author} {\bibfnamefont {V.~N.}\ \bibnamefont {Ciriano-Tejel}}, \bibinfo {author} {\bibfnamefont {D.~F.}\ \bibnamefont {Wise}}, \bibinfo {author} {\bibfnamefont {M.~A.}\ \bibnamefont {Fogarty}}, \bibinfo {author} {\bibfnamefont {T.}~\bibnamefont {Lundberg}}, \bibinfo {author} {\bibfnamefont {C.}~\bibnamefont {Lain\'e}}, \bibinfo {author} {\bibfnamefont {S.}~\bibnamefont {Schaal}}, \bibinfo {author} {\bibfnamefont {F.}~\bibnamefont {Martins}}, \bibinfo {author} {\bibfnamefont {D.~J.}\ \bibnamefont {Ibberson}}, \bibinfo {author} {\bibfnamefont {L.}~\bibnamefont {Hutin}}, \bibinfo {author} {\bibfnamefont {B.}~\bibnamefont {Bertrand}}, \bibinfo {author} {\bibfnamefont {N.}~\bibnamefont {Stelmashenko}}, \bibinfo {author} {\bibfnamefont {J.~W.~A.}\ \bibnamefont {Robinson}}, \bibinfo {author} {\bibfnamefont {L.}~\bibnamefont {Ibberson}}, \bibinfo {author} {\bibfnamefont {A.}~\bibnamefont {Hashim}}, \bibinfo {author}
  {\bibfnamefont {I.}~\bibnamefont {Siddiqi}}, \bibinfo {author} {\bibfnamefont {A.}~\bibnamefont {Lee}}, \bibinfo {author} {\bibfnamefont {M.}~\bibnamefont {Vinet}}, \bibinfo {author} {\bibfnamefont {C.~G.}\ \bibnamefont {Smith}}, \bibinfo {author} {\bibfnamefont {J.~J.~L.}\ \bibnamefont {Morton}},\ and\ \bibinfo {author} {\bibfnamefont {M.~F.}\ \bibnamefont {Gonzalez-Zalba}},\ }\bibfield  {title} {\bibinfo {title} {Fast high-fidelity single-shot readout of spins in silicon using a single-electron box},\ }\href {https://doi.org/10.1103/PhysRevX.13.011023} {\bibfield  {journal} {\bibinfo  {journal} {Phys. Rev. X}\ }\textbf {\bibinfo {volume} {13}},\ \bibinfo {pages} {011023} (\bibinfo {year} {2023})}\BibitemShut {NoStop}%
\bibitem [{\citenamefont {Huang}\ and\ \citenamefont {Hu}(2014)}]{Huang2014}%
  \BibitemOpen
  \bibfield  {author} {\bibinfo {author} {\bibfnamefont {P.}~\bibnamefont {Huang}}\ and\ \bibinfo {author} {\bibfnamefont {X.}~\bibnamefont {Hu}},\ }\bibfield  {title} {\bibinfo {title} {Spin relaxation in a {S}i quantum dot due to spin-valley mixing},\ }\href {https://doi.org/10.1103/PhysRevB.90.235315} {\bibfield  {journal} {\bibinfo  {journal} {Phys. Rev. B}\ }\textbf {\bibinfo {volume} {90}},\ \bibinfo {pages} {235315} (\bibinfo {year} {2014})}\BibitemShut {NoStop}%
\bibitem [{\citenamefont {Tahan}\ and\ \citenamefont {Joynt}(2014)}]{Tahan2014}%
  \BibitemOpen
  \bibfield  {author} {\bibinfo {author} {\bibfnamefont {C.}~\bibnamefont {Tahan}}\ and\ \bibinfo {author} {\bibfnamefont {R.}~\bibnamefont {Joynt}},\ }\bibfield  {title} {\bibinfo {title} {Relaxation of excited spin, orbital, and valley qubit states in ideal silicon quantum dots},\ }\href {https://doi.org/10.1103/PhysRevB.89.075302} {\bibfield  {journal} {\bibinfo  {journal} {Phys. Rev. B}\ }\textbf {\bibinfo {volume} {89}},\ \bibinfo {pages} {075302} (\bibinfo {year} {2014})}\BibitemShut {NoStop}%
\bibitem [{\citenamefont {Yang}\ \emph {et~al.}(2013)\citenamefont {Yang}, \citenamefont {Rossi}, \citenamefont {Ruskov}, \citenamefont {Lai}, \citenamefont {Mohiyaddin}, \citenamefont {Lee}, \citenamefont {Tahan}, \citenamefont {Klimeck}, \citenamefont {Morello},\ and\ \citenamefont {Dzurak}}]{Yang2013}%
  \BibitemOpen
  \bibfield  {author} {\bibinfo {author} {\bibfnamefont {C.~H.}\ \bibnamefont {Yang}}, \bibinfo {author} {\bibfnamefont {A.}~\bibnamefont {Rossi}}, \bibinfo {author} {\bibfnamefont {R.}~\bibnamefont {Ruskov}}, \bibinfo {author} {\bibfnamefont {N.~S.}\ \bibnamefont {Lai}}, \bibinfo {author} {\bibfnamefont {F.~A.}\ \bibnamefont {Mohiyaddin}}, \bibinfo {author} {\bibfnamefont {S.}~\bibnamefont {Lee}}, \bibinfo {author} {\bibfnamefont {C.}~\bibnamefont {Tahan}}, \bibinfo {author} {\bibfnamefont {G.}~\bibnamefont {Klimeck}}, \bibinfo {author} {\bibfnamefont {A.}~\bibnamefont {Morello}},\ and\ \bibinfo {author} {\bibfnamefont {A.~S.}\ \bibnamefont {Dzurak}},\ }\bibfield  {title} {\bibinfo {title} {Spin-valley lifetimes in a silicon quantum dot with tunable valley splitting},\ }\bibfield  {journal} {\bibinfo  {journal} {Nature Communications}\ }\textbf {\bibinfo {volume} {4}},\ \href {https://doi.org/10.1038/ncomms3069} {10.1038/ncomms3069} (\bibinfo {year} {2013})\BibitemShut {NoStop}%
\bibitem [{\citenamefont {Ciriano-Tejel}\ \emph {et~al.}(2021)\citenamefont {Ciriano-Tejel}, \citenamefont {Fogarty}, \citenamefont {Schaal}, \citenamefont {Hutin}, \citenamefont {Bertrand}, \citenamefont {Ibberson}, \citenamefont {Gonzalez-Zalba}, \citenamefont {Li}, \citenamefont {Niquet}, \citenamefont {Vinet},\ and\ \citenamefont {Morton}}]{Virginia2021}%
  \BibitemOpen
  \bibfield  {author} {\bibinfo {author} {\bibfnamefont {V.~N.}\ \bibnamefont {Ciriano-Tejel}}, \bibinfo {author} {\bibfnamefont {M.~A.}\ \bibnamefont {Fogarty}}, \bibinfo {author} {\bibfnamefont {S.}~\bibnamefont {Schaal}}, \bibinfo {author} {\bibfnamefont {L.}~\bibnamefont {Hutin}}, \bibinfo {author} {\bibfnamefont {B.}~\bibnamefont {Bertrand}}, \bibinfo {author} {\bibfnamefont {L.}~\bibnamefont {Ibberson}}, \bibinfo {author} {\bibfnamefont {M.~F.}\ \bibnamefont {Gonzalez-Zalba}}, \bibinfo {author} {\bibfnamefont {J.}~\bibnamefont {Li}}, \bibinfo {author} {\bibfnamefont {Y.-M.}\ \bibnamefont {Niquet}}, \bibinfo {author} {\bibfnamefont {M.}~\bibnamefont {Vinet}},\ and\ \bibinfo {author} {\bibfnamefont {J.~J.}\ \bibnamefont {Morton}},\ }\bibfield  {title} {\bibinfo {title} {Spin readout of a {CMOS} quantum dot by gate reflectometry and spin-dependent tunneling},\ }\href {https://doi.org/10.1103/PRXQuantum.2.010353} {\bibfield  {journal} {\bibinfo  {journal} {PRX Quantum}\ }\textbf {\bibinfo {volume} {2}},\
  \bibinfo {pages} {010353} (\bibinfo {year} {2021})}\BibitemShut {NoStop}%
\bibitem [{\citenamefont {Petit}\ \emph {et~al.}(2018)\citenamefont {Petit}, \citenamefont {Boter}, \citenamefont {Eenink}, \citenamefont {Droulers}, \citenamefont {Tagliaferri}, \citenamefont {Li}, \citenamefont {Franke}, \citenamefont {Singh}, \citenamefont {Clarke}, \citenamefont {Schouten}, \citenamefont {Dobrovitski}, \citenamefont {Vandersypen},\ and\ \citenamefont {Veldhorst}}]{Petit2018}%
  \BibitemOpen
  \bibfield  {author} {\bibinfo {author} {\bibfnamefont {L.}~\bibnamefont {Petit}}, \bibinfo {author} {\bibfnamefont {J.~M.}\ \bibnamefont {Boter}}, \bibinfo {author} {\bibfnamefont {H.~G.~J.}\ \bibnamefont {Eenink}}, \bibinfo {author} {\bibfnamefont {G.}~\bibnamefont {Droulers}}, \bibinfo {author} {\bibfnamefont {M.~L.~V.}\ \bibnamefont {Tagliaferri}}, \bibinfo {author} {\bibfnamefont {R.}~\bibnamefont {Li}}, \bibinfo {author} {\bibfnamefont {D.~P.}\ \bibnamefont {Franke}}, \bibinfo {author} {\bibfnamefont {K.~J.}\ \bibnamefont {Singh}}, \bibinfo {author} {\bibfnamefont {J.~S.}\ \bibnamefont {Clarke}}, \bibinfo {author} {\bibfnamefont {R.~N.}\ \bibnamefont {Schouten}}, \bibinfo {author} {\bibfnamefont {V.~V.}\ \bibnamefont {Dobrovitski}}, \bibinfo {author} {\bibfnamefont {L.~M.~K.}\ \bibnamefont {Vandersypen}},\ and\ \bibinfo {author} {\bibfnamefont {M.}~\bibnamefont {Veldhorst}},\ }\bibfield  {title} {\bibinfo {title} {Spin lifetime and charge noise in hot silicon quantum dot qubits},\ }\href
  {https://doi.org/10.1103/PhysRevLett.121.076801} {\bibfield  {journal} {\bibinfo  {journal} {Phys. Rev. Lett.}\ }\textbf {\bibinfo {volume} {121}},\ \bibinfo {pages} {076801} (\bibinfo {year} {2018})}\BibitemShut {NoStop}%
\bibitem [{\citenamefont {Veldhorst}\ \emph {et~al.}(2014)\citenamefont {Veldhorst}, \citenamefont {Hwang}, \citenamefont {Yang}, \citenamefont {Leenstra}, \citenamefont {Ronde}, \citenamefont {Dehollain}, \citenamefont {Muhonen}, \citenamefont {Hudson}, \citenamefont {Itoh}, \citenamefont {Morello},\ and\ \citenamefont {Dzurak}}]{Veldhorst2014}%
  \BibitemOpen
  \bibfield  {author} {\bibinfo {author} {\bibfnamefont {M.}~\bibnamefont {Veldhorst}}, \bibinfo {author} {\bibfnamefont {J.~C.}\ \bibnamefont {Hwang}}, \bibinfo {author} {\bibfnamefont {C.~H.}\ \bibnamefont {Yang}}, \bibinfo {author} {\bibfnamefont {A.~W.}\ \bibnamefont {Leenstra}}, \bibinfo {author} {\bibfnamefont {B.~D.}\ \bibnamefont {Ronde}}, \bibinfo {author} {\bibfnamefont {J.~P.}\ \bibnamefont {Dehollain}}, \bibinfo {author} {\bibfnamefont {J.~T.}\ \bibnamefont {Muhonen}}, \bibinfo {author} {\bibfnamefont {F.~E.}\ \bibnamefont {Hudson}}, \bibinfo {author} {\bibfnamefont {K.~M.}\ \bibnamefont {Itoh}}, \bibinfo {author} {\bibfnamefont {A.}~\bibnamefont {Morello}},\ and\ \bibinfo {author} {\bibfnamefont {A.~S.}\ \bibnamefont {Dzurak}},\ }\bibfield  {title} {\bibinfo {title} {An addressable quantum dot qubit with fault-tolerant control-fidelity},\ }\href {https://doi.org/10.1038/nnano.2014.216} {\bibfield  {journal} {\bibinfo  {journal} {Nature Nanotechnology}\ }\textbf {\bibinfo {volume} {9}},\ \bibinfo
  {pages} {981} (\bibinfo {year} {2014})}\BibitemShut {NoStop}%
\bibitem [{\citenamefont {de~Fuentes}\ \emph {et~al.}(2025)\citenamefont {de~Fuentes}, \citenamefont {Raymenants}, \citenamefont {Undseth}, \citenamefont {Pietx-Casas}, \citenamefont {Mądzik}, \citenamefont {de~Snoo}, \citenamefont {Amitonov}, \citenamefont {Tryputen}, \citenamefont {Schmitz}, \citenamefont {Matsuura}, \citenamefont {Scappucci},\ and\ \citenamefont {Vandersypen}}]{Irene2025}%
  \BibitemOpen
  \bibfield  {author} {\bibinfo {author} {\bibfnamefont {I.~F.}\ \bibnamefont {de~Fuentes}}, \bibinfo {author} {\bibfnamefont {E.}~\bibnamefont {Raymenants}}, \bibinfo {author} {\bibfnamefont {B.}~\bibnamefont {Undseth}}, \bibinfo {author} {\bibfnamefont {O.}~\bibnamefont {Pietx-Casas}}, \bibinfo {author} {\bibfnamefont {S.~P.~M.}\ \bibnamefont {Mądzik}}, \bibinfo {author} {\bibfnamefont {S.~L.}\ \bibnamefont {de~Snoo}}, \bibinfo {author} {\bibfnamefont {S.~V.}\ \bibnamefont {Amitonov}}, \bibinfo {author} {\bibfnamefont {L.}~\bibnamefont {Tryputen}}, \bibinfo {author} {\bibfnamefont {A.~T.}\ \bibnamefont {Schmitz}}, \bibinfo {author} {\bibfnamefont {A.~Y.}\ \bibnamefont {Matsuura}}, \bibinfo {author} {\bibfnamefont {G.}~\bibnamefont {Scappucci}},\ and\ \bibinfo {author} {\bibfnamefont {L.~M.~K.}\ \bibnamefont {Vandersypen}},\ }\href {https://arxiv.org/abs/2505.19200} {\bibinfo {title} {Running a six-qubit quantum circuit on a silicon spin qubit array}} (\bibinfo {year} {2025}),\BibitemShut {NoStop}%
\bibitem [{\citenamefont {Barthel}\ \emph {et~al.}(2009)\citenamefont {Barthel}, \citenamefont {Reilly}, \citenamefont {Marcus}, \citenamefont {Hanson},\ and\ \citenamefont {Gossard}}]{Barthel2009}%
  \BibitemOpen
  \bibfield  {author} {\bibinfo {author} {\bibfnamefont {C.}~\bibnamefont {Barthel}}, \bibinfo {author} {\bibfnamefont {D.~J.}\ \bibnamefont {Reilly}}, \bibinfo {author} {\bibfnamefont {C.~M.}\ \bibnamefont {Marcus}}, \bibinfo {author} {\bibfnamefont {M.~P.}\ \bibnamefont {Hanson}},\ and\ \bibinfo {author} {\bibfnamefont {A.~C.}\ \bibnamefont {Gossard}},\ }\bibfield  {title} {\bibinfo {title} {Rapid single-shot measurement of a singlet-triplet qubit},\ }\href {https://doi.org/10.1103/PhysRevLett.103.160503} {\bibfield  {journal} {\bibinfo  {journal} {Phys. Rev. Lett.}\ }\textbf {\bibinfo {volume} {103}},\ \bibinfo {pages} {160503} (\bibinfo {year} {2009})}\BibitemShut {NoStop}%
\bibitem [{\citenamefont {van~den Burg}\ and\ \citenamefont {Williams}(2022)}]{burg2022evaluationchangepointdetection}%
  \BibitemOpen
  \bibfield  {author} {\bibinfo {author} {\bibfnamefont {G.~J.~J.}\ \bibnamefont {van~den Burg}}\ and\ \bibinfo {author} {\bibfnamefont {C.~K.~I.}\ \bibnamefont {Williams}},\ }\href {https://arxiv.org/abs/2003.06222} {\bibinfo {title} {An evaluation of change point detection algorithms}} (\bibinfo {year} {2022}),\ \BibitemShut {NoStop}%
  


\end{thebibliography}

%\bibliographystyle{unsrt}
%\input{main_text.bbl}

%
\makeatother        % <-- add this line

\end{document}